\documentclass[a4paper,11pt]{article}
\pdfoutput=1 
\usepackage{jheppub}
\usepackage[T1]{fontenc} 
\def\be{\begin{equation}}
\def\ee{\end{equation}}
\def\bea{\begin{array}}
\def\eea{\end{array}}
\def\beqa{\begin{eqnarray}}
\def\eeqa{\end{eqnarray}}
\def\beqas{\begin{eqnarray*}}
\def\eeqas{\end{eqnarray*}}

\def\bp{\begin{picture}}
\def\ep{\end{picture}}
\def\bc{\begin{center}}
\def\ec{\end{center}}
\def\bfig{\begin{figure}}
\def\efig{\end{figure}}

\def\bit{\begin{itemize}}
\def\eit{\end{itemize}}
\def\nn{\nonumber}
\def\f{\frac}

\def\[{\left[}
\def\]{\right]}
\def\({\left(}
\def\){\right)}

\def\..{\left.}
\def\.{\right.}
\def\tl{\tilde}
\def\ra{\rightarrow}
\def\la{\leftarrow}

\def\tm{\times}

\def\da{\dagger}

\def\la{\lambda}

\def\al{\alpha}

\def\ka{\kappa}

\def\si{\sigma}
\def\ep{\epsilon}

\def\ga{\gamma}
\def\pa{\partial}
\def\pr{\prime}

\title{\boldmath Type-II neutrino seesaw mechanism extension of NMSSM from SUSY breaking mechanisms}
\author[a]{Zhuang Li,}
\author[a,1]{Fei Wang\note{Corresponding author.}}
\affiliation[a]{School of Physics, Zhengzhou University,450000,ZhengZhou, P.R.China}
\emailAdd{feiwang@zzu.edu.cn}

\abstract{We propose to accommodate economically the type-II neutrino seesaw mechanism in (G)NMSSM from GMSB and AMSB, respectively. The heavy triplets within neutrino seesaw mechanism are identified to be the messengers. Therefore, the $\mu$-problem, the neutrino mass generation, LFV as well the soft SUSY breaking parameters can be economically combined in a non-trivial way. General features of such extensions are discussed. The type-II neutrino seesaw-specific interactions can give additional Yukawa deflection contributions to the soft SUSY breaking parameters of NMSSM, which are indispensable to realize successful EWSB and accommodate the 125 GeV Higgs. Relevant numerical results, including the constraints of dark matter and possible LFV processes $l_i\ra l_j \ga$ etc, are also given. We find that our economical type-II neutrino seesaw mechanism extension of NMSSM from AMSB or GMSB can lead to realistic low energy NMSSM spectrum, both admitting the 125 GeV Higgs as the lightest CP-even scalar. The possibility of the 125 GeV Higgs being the next-to-lightest CP-even scalar in GMSB-type scenario is ruled out by the constraints from EWSB, collider and precision measurements. The possibility of the 125 GeV Higgs being the next-to-lightest CP-even scalar in AMSB-type scenario is ruled out by dark matter direct detection experiments. Possible constraints from LFV processes $l_i\ra l_j \ga$ can give an upper bound for the messenger scale. }
\begin{document}
\maketitle
\flushbottom

\section{Introduction}

TeV scale supersymmetry(SUSY) is one of the most promising candidates for new physics beyond the Standard Model(SM). It can prevent the Higgs boson mass from acquiring dangerous quadratic divergence corrections, realize successful gauge coupling unification and provide viable dark matter(DM) candidates, such as the lightest neutralino assuming exact R-parity. Besides, the discovered 125 GeV Higgs~\cite{ATLAS:higgs,CMS:higgs} lies miraculously in the narrow 115-135 GeV $'window'$ predicted by MSSM, which can also be seen as a triumph of low scale SUSY. However, low energy SUSY confronts many challenges from LHC experiments, the foremost of which is the null search results of superpartners at LHC. Recent analyses based on Run 2 of 13 TeV LHC and $36 fb^{-1}$ of integrated luminosity constrain the gluino mass $m_{\tl{g}}$ to lie above 2 TeV~\cite{CMSSM1} and the top squark mass $m_{\tl{t}_1}$ to lie above 1 TeV~\cite{CMSSM2} in some simplified models. In addition, the $\mu$ problem in MSSM needs an explanation.

One of the major unresolved problems of particle physics now is the nature of tiny neutrino masses, which were discovered by neutrino oscillation experiments. It is known that Weinberg's effective dimension-5 operator is the lowest one which can generate tiny Majorana neutrino masses. Such an operator can be  ultraviolet(UV)-completed to obtain three types of tree-level seesaw mechanism: type I seesaw~\cite{type I}, involving the exchange of right-handed neutrinos; type II seesaw~\cite{type II}, involving the exchange of scalar triplet; type III~\cite{type III}, involving the exchange of fermion triplet. If the SUSY framework is indeed the new physics beyond the SM, it should accommodate proper neutrino mass generation mechanisms. The seesaw mechanism extensions of low energy SUSY~\cite{typeI+MSSM}, which can provide typical unified frameworks to solve all the remaining puzzles of SM together, are well motivated theoretically.

However, simple seesaw mechanism extensions of MSSM still inherit the main difficulties of MSSM. The foremost one is the $\mu$-problem, which is in general unsolved in such extensions. Besides, to accommodate the 125 GeV Higgs, unnaturally heavy stop masses $m_{\tl{t}}\gtrsim 5$ TeV are necessary unless large trilinear coupling $A_t$ is present, which on the other hand may result in color breaking minimum for the scalar potential~\cite{VEV:color}. Next-to-minimal supersymmetric standard model(NMSSM)~\cite{NMSSM} is the simplest gauge singlet extension of MSSM, which can elegantly solve the $\mu$-problem in MSSM by generating an effective $\mu$-term after the singlet scalar acquires a vacuum expectation value (VEV). Furthermore, due to possible new tree level contributions to the Higgs mass, NMSSM can easily accommodate the 125 GeV Higgs boson without the needs of very large $A_t$ for light stops, ameliorating the color breaking minimum problem. Therefore, the seesaw mechanism extensions of NMSSM can evade most of the difficulties that bother the seesaw mechanism extensions of MSSM. Attracting as the seesaw mechanism extensions of NMSSM are, there are too many free parameters in such  low energy SUSY models. To preserve their prediction power, we need to refer to their UV completion. It is known that the low energy SUSY spectrum can be totally determined by the SUSY breaking mechanism, which can predict the low energy parameters by very few UV inputs. So it is desirable to combine the seesaw mechanism extensions of (N)MSSM with the SUSY breaking mechanisms and survey which SUSY breaking mechanism can give the favored low energy spectrum.

Depending on the way  the visible sector $'feels'$ the SUSY breaking effects in the hidden sector, the SUSY breaking mechanisms can be classified into gravity mediation~\cite{SUGRA}, gauge mediation~\cite{GMSB}(GMSB), anomaly mediation~\cite{AMSB}(AMSB) scenarios, etc. Both GMSB and AMSB are calculable, predictive, and phenomenologically distinctive. Especially, they will not cause flavor and CP problems that bothers gravity mediation models. However, GMSB realization of MSSM can hardly explain the 125 GeV Higgs with TeV scale soft SUSY breaking parameters because of the vanishing trilinear terms at the messenger scale. Although non-vanishing $A_t$ at the messenger scale can be obtained in GMSB with additional messenger-matter interactions~\cite{chacko,shih,gmsb-mm}, it is rather ad hoc to include such interactions in the superpotential. So it is interesting to see if certain types of messenger-matter interactions can arise naturally in an UV-completed model. Yukawa mediation contributions from messenger-matter interactions can also possibly be present~\cite{MM:DAMSB} in deflected AMSB~\cite{dAMSB,okada}, which can elegantly solve the tachyonic slepton problem of minimal AMSB through the deflection of the renormalization group equation (RGE) trajectory ~\cite{deflect:RGE-invariance}.

It is fairly straightforward to accommodate SUSY breaking mechanism in the neutrino-seesaw extended MSSM, for example, by introducing additional messenger sector in GMSB or (deflected) AMSB. However, to introduce as few new inputs as possible, it is more predictive and economical to identify the messengers with the heavy fields that are integrated out in the neutrino-seesaw mechanism. In such predictive models, the neutrino mass generation, lepton-flavor-violation(LFV) as well soft SUSY breaking parameters can be related together. Besides, additional Yukawa couplings involving the heavy fields (in neutrino seesaw mechanism), which also act as the messengers, can be naturally present. Such messenger-matter type interactions can possibly give large contributions to trilinear $A_t$ term in GMSB (or deflected AMSB), which will play an important rule in obtaining the 125 GeV Higgs with TeV scale soft SUSY breaking parameters.

As noted previously, even though it is very predictive and well motivated to combine neutrino seesaw mechanism with SUSY breaking mechanism~\cite{typeII+GMSB,typeII+AMSB} in an economical way for MSSM, the difficulties of MSSM mentioned previously, especially the $\mu$-problem, are in general not solved, making it interesting to turn instead to such realizations of NMSSM. As the case of MSSM, it is in general straightforward to accommodate SUSY breaking mechanism in the neutrino-seesaw extended NMSSM by introducing an additional messenger sector in GMSB(dAMSB). Nevertheless, it is still interesting to see if the neutrino mass generation, LFV, the generation of $\mu$-term as well the soft SUSY breaking parameters can be combined in a non-trivial economical way by identifying the messengers with the heavy fields. Such an economical realization of Type I seesaw extension of NMSSM from GMSB, which introduce only gauge singlet neutrino superfields, can hardly generate soft SUSY breaking parameters other than the left-handed sleptons and right-handed sneutrinos without additional non-singlet messengers\footnote{The superpotential of type-I seesaw extension of NMSSM \cite{typeI+NMSSM}can naively be embedded economically in Yukawa mediation with
\beqas
W_{\rm Type~I}\supseteq y_{ij}^N L_i H_u N_j+ X N_j^2~,
\eeqas
which, however, can not generate realistic spectrum. Here $N_j$ the right-handed neutrino superfields and $X$ the SUSY breaking spurion superfield with its VEV $\langle X\rangle=M+\theta^2 F_X$. The lowest component VEV of $X$ can determine the $N_j$ thresholds.
 The inverse seesaw extension of NMSSM\cite{ISS1,ISS2,ISS5,ISS3,ISS4} can be written as
  \beqas
 W_{\rm Inverse}\supseteq y_{ij}^N L_i H_u N_j+ \tl{\la} S N_j N_0+ \mu_X N_0 {N}_0~.
  \eeqas
with the presence of a very small lepton number violating parameter $\mu_X \sim {\rm eV}$ that is responsible for the smallness of the light neutrino masses. $N_0$ is the additional gauge singlet field. This extension does not have a similar economical GMSB embedding and can be embedded in ordinary realization of GMSB with an additional messenger sector or (dAMSB). A successful realization can be seen in~\cite{1405.5478}.}.
   Similar extension in AMSB, however, cannot lead to positive squared masses for right-handed sleptons. The economical realization of Type II neutrino seesaw extension of NMSSM, on the other hand, can generate realistic soft SUSY parameters without the need of an additional messenger sector other than the heavy fields present in the seesaw mechanism. We will discuss the realization of NMSSM through the economical combination of type-II neutrino seesaw mechanism with GMSB and deflected AMSB, respectively.

   This paper is organized as follows. In Sec \ref{type II}, we discuss the type-II neutrino seesaw mechanism in SUSY. In Sec \ref{GMSB} and Sec \ref{AMSB}, we discuss the economical realization of type-II seesaw mechanism extension of NMSSM from GMSB and AMSB, respectively. The soft SUSY breaking parameters are given and numerical results for each scenarios are studied. Sec \ref{conclusions} contains our conclusions.

\section{\label{type II} Type-II neutrino seesaw mechanism in SUSY}
In the ordinary type-II seesaw mechanism~\cite{type II}, the Lagrangian contains the coupling between the scalar triplet to the Higgs doublet $H$ as well as the Yukawa interaction between the $SU(2)_L$ doublet leptons to a very heavy $SU(2)_L$ triplet scalar with lepton number $L=-2$ and mass $M_\Delta$
\beqa
\label{type II:SM}
{\cal L}\supset -M_\Delta^2 |\Delta_L|^2+ y_{ij}^\nu L_{L;i}^T C \Delta_L L_{L;j}+ A_T H^T \Delta_L H~.
\eeqa
 The third term, which contains a trilinear scalar coupling mass parameter $A_T$, plays a key role in determining the minimum of the full scalar potential so as to give a tiny vacuum expectation value(VEV) of $\Delta_L$. Such a tiny VEV can in turn induce a Majorana mass for left-handed neutrinos
 \beqa
 m_\nu\approx y_{ij}^\nu \f{A_T v^2}{M^2_\Delta}\sim 0.1 {\rm eV}~,
 \eeqa
 with $v \approx 246 {\rm GeV}$. For $y_{ij}^\nu \sim {\cal O}(1)$, $M_\Delta\sim 10^{14}$ GeV in the case $A_T\sim M_\Delta$ and $M_\Delta\sim 10^8$ GeV in the case $A_T\sim v_{EW}$.

The type-II neutrino seesaw mechanism extension of MSSM is non-trivial.  There are two $SU(2)_L$ Higgs doublets in the MSSM, so the type-II neutrino seesaw mechanism extension of MSSM can be seen as a special case of type-II neutrino seesaw extension of two Higgs doublet model, which contains interactions between both scalar Higgs doublets to the heavy scalar triplet. We can further extend the type-II seesaw mechanism to NMSSM by including the singlet sector.

 We need to introduce vector-like $SU(2)_L$ triplet superfields with $U(1)_Y$ quantum number $Y=\pm 2$ in the superpotential
\beqa
\label{GNMSSM+TypeII}
W_1&\supseteq& W_{\rm NMSSM}+y^L_{ij} L_j L_i \Delta_T +y^d_\Delta \Delta_T H_d H_d+m_T\overline{\Delta}_T \Delta_T+y^u_\Delta \overline{\Delta}_T H_u H_u~,
\eeqa
with general NMSSM superpotential\footnote{For $Z_3$-invariant NMSSM, the $y^d_\Delta \Delta_T H_d H_d$ term can be forbidden by proper $Z_3$ charge assignments. }
\beqa
W_{\rm NMSSM}=W_{\rm MSSM/\mu} +\la S H_u H_d+\f{\ka}{3}S^3+\xi_S S+\cdots~.
\eeqa
The parameter $m_T$, which is a free parameter in equation (\ref{GNMSSM+TypeII}), will be determined by the spurion VEVs in GMSB (or deflected AMSB) if the triplets can act as components of the messengers.\footnote{We should note that it is consistent to generate $m_T$ also by the VEV of $S$ for tiny coupling $y_{ij}^\nu$ in NMSSM. However, large fine tuning will be needed in general because of large effective $\mu$.}
From the superpotential (\ref{GNMSSM+TypeII}), we can obtain the F-terms of the triplets
\beqa
F_{\Delta_T}&=&\f{\pa W_1}{\pa \Delta_T}= y^L_{ij}L_i L_j+y_\Delta^d H_d H_d+m_T \overline{\Delta}_T~,\nn\\
F_{\overline{\Delta}_T}&=&\f{\pa W}{\pa \overline{\Delta}_T}= y_\Delta^u H_u H_u+m_T {\Delta}_T~.
\eeqa
We require the SUSY to be unbroken at the triplet scale $m_T$. So the F-flat conditions $F_{\Delta_T}=F_{\overline{\Delta}_T}=0$ can give
\beqa
\label{T-flat}
\langle\overline{\Delta}_T\rangle=-y_\Delta^d \f{v_d^2}{m_T}~,~~~~
\langle {\Delta}_T\rangle=-y_\Delta^u \f{v_u^2}{m_T}~.
\eeqa

The neutrinos will acquire tiny Majorana masses through the type-II seesaw mechanism
\beqa\label{SUSYSeesaw:TypeII}
 \langle m_\nu \rangle
=-y^L_{ij} y_\Delta^u \f{v_u^2}{m_T}.
\eeqa
This result can be understood to arise from the scalar potential
\beqa
\label{potential}
V&\supset&\left|y^L_{ij}L_i L_j+y_\Delta^d H_d H_d+m_T \overline{\Delta}_T\right|^2+\left|\f{}{}y_\Delta^u H_u H_u+m_T {\Delta}_T\right|^2+
\left|\la S H_u+2y_\Delta^d\Delta_T H_d\right|\nn\\
&+&\left|\la S H_d+2y_\Delta^u\overline{\Delta}_T H_u\right| +\cdots~.
\eeqa
The $m_T y_\Delta^u H_u^* H_u^* {\Delta}_T$ term plays the role of the third term in (\ref{type II:SM}).

 We should note that F-terms $F_{H_u}$ for $H_u$ and $F_{H_d}$ for $H_d$ can not vanish for solutions in eqn (\ref{T-flat}) with non-negligible $\mu$ term. Therefore, tiny SUSY breaking effects of order $|F_{H_u}|^2+|F_{H_d}|^2\sim \mu^2 v^2$ will appear. In fact, the minimum conditions for $H_u$ and $H_d$ should also involve the soft SUSY breaking terms.

 From the potential eqn.(\ref{potential}), we can see that the term involving $\mu$ with \beqa\la \langle S \rangle 2y_\Delta^d \Delta_T H_d H_u^*=2y_\Delta^d \mu \Delta_T H_d H_u^*~,\eeqa
also gives a subleading contribution to neutrino masses. Besides, there is an alternative contribution to neutrino masses from the trilinear soft SUSY breaking term
\beqa
-{\cal L}\supseteq A_{H_d H_d \Delta_T}y^d_\Delta \Delta_T H_d H_d+\cdots~,
\eeqa
which will be generated after SUSY breaking. From the minimum conditions of the total scalar potential, including the soft SUSY breaking terms, the triplet VEV can be approximately given by
\beqa
\langle {\Delta}_T\rangle\approx -y_\Delta^u \f{v_u^2}{m_T}-y_\Delta^d\f{A_{H_dH_d\Delta_T}v_d^2}{m_T^2}-y_\Delta^d\f{2 \mu v_d v_u}{m_T^2}~.
\eeqa
So the resulting neutrino masses are given by
\beqa
\label{neutrino:full}
\(m_\nu\)_{ij}= -y^L_{ij}\[ y_\Delta^u \f{v_u^2}{m_T}+y_\Delta^d\f{A_{H_dH_d\Delta_T}v_d^2}{m_T^2}+y_\Delta^d\f{2 \mu \tan\beta v_d^2}{m_T^2}\] \lesssim ~0.1 {\rm eV}.
\eeqa
The three terms can be destructive if $A_{H_dH_d\Delta_T}$ or $\mu$ is negative. Besides, if $|A_{H_dH_d\Delta_T}|\gtrsim m_T$ for negative $A_{H_dH_d\Delta_T}$ or similarly for $\mu$, tiny neutrino masses can be generated by fine tuning even if either terms in eqn (\ref{neutrino:full}) are not very small.

Such a type-II neutrino seesaw mechanism extension of (N)MSSM can be nontrivially embedded into SUSY breaking mechanisms.
In this paper, the messenger threshold can be identified to be the heavy triplet scalar threshold in type-II seesaw mechanism. This possibility provide an economic unified framework to taking into account both SUSY extension and neutrino masses. So $m_T$ is always much larger than the $A_{H_dH_d\Delta_T}$, which lies typically at the soft SUSY breaking scale. Successful EWSB requires $\mu$ to lie at the soft SUSY breaking scale. Therefore, the second and third terms in eqn (\ref{neutrino:full}) are always subleading unless the messenger scale is very low.

The messenger threshold, which is just the heavy scalar triplet scale in type-II neutrino seesaw mechanism, can possibly be constrained by the lepton flavor violation(LFV) processes, such as $l_i\ra l_j \ga$. Detailed discussions on LFV constraints to SUSY seesaw models can be found in ~\cite{hep-ph:0103065,hep-ph:0106245,LFV:1207.6635}. Especially, the LFV related discussions in scenario within which the triplet in type-II neutrino seesaw mechanism also account for the soft SUSY breaking masses had been discussed in~\cite{typeII+GMSB,hep-ph:0607298}.

The branch ratio $l_i\ra l_j \ga$ can be generally written as~\cite{LFV:ega}
\beqa
Br(l_i\ra l_j \ga)=\f{48\pi^3\al_{e}}{G_F}\(|A_L^{ij}|^2+|A_R^{ij}|^2\)Br(l_i\ra l_j \nu_i\bar{\nu}_j)~,
\eeqa
where
\beqa
A_L^{ij}\approx \f{\(m^2_{\tl{L}}\)_{ij}}{m_{SUSY}^4}~,~~A_R^{ij}\approx \f{\(m^2_{\tl{E}_L^c}\)_{ij}}{m_{SUSY}^4}~,
\eeqa
with $m^2_{\tl{L}}$ and $m^2_{\tl{E}_L^c}$ are the doublet and singlet slepton soft mass matrices, respectively. $m_{SUSY}$ is the typical SUSY mass scale.  These estimations depend on the assumptions that (I) chargino/neutralino masses are similar to slepton masses and (II) left-right
flavor mixing induced by trilinear terms is negligible. As noted in ~\cite{LFV:1207.6635}, although the assumption is not valid when large values of trilinear terms are considered, the above estimates
can nevertheless be used to illustrate the dependence of the BRs on the low-energy neutrino parameters.

To avoid severe difficulties from SUSY flavor constraints, the soft sfermion masses (including the slepton masses) are universal at high energy input scale $M_U$. The RGEs of the slepton soft terms, which contain non-diagonal contributions from neutrino-seesaw specific interactions, can possibly induce off-diagonal soft terms to slepton mass matrices. These contributions are decoupled at the characteristic scale of the heavy mediators $m_T$. However, it is interesting to note that in our subsequent discussions with gauge mediation and (deflected) anomaly mediation, the trilinear couplings for slepton Yukawa (with $A_{E ij}\approx 0$) and slepton masses are universal at the (input) messenger scale, which also act as the heavy triplet mediator scale. In the basis where the lepton Yukawa couplings are diagonal, all the LFV effects are encoded in the coupling $Y_{ij}^L$. From the RGE of the soft masses~\cite{hep-ph:0103065}, one can obtain the leading-log approximation~\cite{LFV:1207.6635,hep-ph:0103065} for the off-diagonal soft terms at low energy
\beqa
\(m^2_{\tl{L}}\)_{ij}&\approx& -\f{6}{8\pi^2} \(3 m^2_{\tl{L}_{L}}\) \[Y^{L\da}_{ik} Y_{kj}^L\]\log\(\f{M_{U}}{m_T}\)~,\nn\\
\(m^2_{\tl{E}_L^c}\)_{ij}&\approx& 0~,~~~A_{E ij}\approx 0~,
\eeqa
with
\beqa
Y^{L\da}_{ik} Y_{kj}^L=\(\f{m_T}{y_\Delta^u v_u^2}\)^2\[U(m_\nu^{diag})^2 U^\da\]_{ij}.
\eeqa
Here $U$ is the PMNS lepton mixing matrix
\beqa
U=\left(\bea{ccc}
c_{12}c_{13}&s_{12}c_{13}&s_{13}e^{-i\delta}\\
-s_{12}c_{23}-c_{12}s_{23}s_{13}e^{i\delta}&c_{12}c_{23}-s_{12}s_{23}s_{13}e^{i\delta}&s_{23}c_{13}\\
s_{12}s_{23}-c_{12}c_{23}s_{13}e^{i\delta}& -c_{12}s_{23} -s_{12}c_{23}s_{13}e^{i\delta}& c_{23}c_{13}
\eea\right)\cdot\left(\bea{ccc}1&&\\&e^{i\phi_1}&\\&&e^{i\phi_2}\eea\right),
\eeqa
with $s_{ij}\equiv \sin\theta_{ij}$, $c_{ij}\equiv\cos\theta_{ij}$ for the three mixing angles $\theta_{12}$, $\theta_{23}$ and $\theta_{13}$, respectively.
 So the BRs for rare lepton decays $l_i\ra l_j \ga$, which are roughly given by
\beqa
Br(l_i\ra l_j \ga)\approx \al_e^3 m_{l_i}^5 \f{\|(m^2_{\tl{L}})_{ij}\|^2}{m_{SUSY}^8}\tan^2\beta~
\propto \[U(m_\nu^{diag})^2 U^\da\]\log\(\f{M_{U}}{m_T}\),
\eeqa
will not receive large enhancement by the log factor with $m_T\sim M_U$ in the leading-log approximation unless the sub-leading terms are sizeable. Therefore, unlike neutrino seesaw mechanism extension of SUGRA-type mediation mechanism, in which the universal soft parameter inputs are adopted at GUT scale with $\log (M_U/m_T)\gg 1$, the BRs of $l_i\ra l_j \ga$ will give important but not too stringent constrains on the seesaw scale in our GMSB and AMSB type scenarios, in which the triplet mediator scale is identified with the messenger scale. Predictions for other LFV processes with best-fit values for the neutrino parameters can be seen in ~\cite{hep-ph:0607298}.

\section{\label{GMSB}Type-II neutrino seesaw mechanism extension of NMSSM from GMSB}
It is known that additional settings are needed to solve the $\mu/B\mu$ problem in ordinary GMSB realization of MSSM. Although such a problem can be naturally solved in NMSSM, additional Yukawa structures for superfield $S$ are needed in GMSB because the soft parameters involving only the singlet $S$ can not receive any gauge mediation contributions. Besides, to accommodate the 125 GeV Higgs in MSSM, TeV scale stop masses with near-maximal stop mixing are necessary~\cite{1112.3068}. For ${\cal O}(10)$ TeV stops with small $A_t$, although still possible to interpret the 125 GeV Higgs, exacerbate the $'little ~hierarchy'$ problem arising from the large mass gap between the measured value of the weak scale and the sparticle mass scale. As ordinary GMSB predicts vanishing $A_t$ at the messenger scale, it necessitates the introduction of additional large Yukawa deflection contributions from messenger-matter interactions if we would like to reduce the fine tuning involved. NMSSM does not need too large $A_t$ for light stops to interpret the 125 GeV  Higgs because of additional tree-level contributions, possibly avoiding the color breaking minimum problem of MSSM. An mildly large $A_t$ can, however, lead to reduced electroweak fine-tuning~\cite{rnaturalsusy}(EWFT) even with TeV scale stops.
 In our economical realization of the type-II neutrino seesaw mechanism extension of NMSSM from GMSB, the Higgs sector can participate in new interactions involving the triplets, which leads to additional non-vanishing Yukawa mediation contributions to trilinear couplings $A_t$ at the messenger scale, possibly reducing the EWFT involved. Besides, additional Yukawa couplings involving $S$ and heavy fields can be naturally introduced, which will also give Yukawa mediation contributions to soft SUSY breaking parameters involving the gauge singlet $S$, making spontaneously symmetry breaking(SSB) possible to give correct range of the $\mu$ value.
Therefore, our predictive type-II neutrino seesaw mechanism extension of NMSSM from GMSB, which can combine the solution to the $\mu$ problem, the neutrino mass generation, LFV, soft masses and EWSB, is very interesting.

\subsection{Theoretical setting of the model}
In GMSB, the VEV of spurion $X$ is given by
\beqa
\langle X\rangle=M+\theta^2 F_X~.
\eeqa
As emphasized in ~\cite{NMSSM}, successful electroweak symmetry breaking(EWSB) in NMSSM necessitates  non-vanishing soft SUSY masses for $S$ and $A_\ka$. As the soft mass of the gauge singlet $S$ receives no contributions from ordinary GMSB, additional Yukawa mediation contributions should be included. It was noted in ~\cite{0706.3873} that double species of messengers are needed to avoid possible mixing between the spurion $X$ and the gauge singlet $S$ if we couple the messengers to $S$. As the $SU(2)_L$ triplet superfields with $SU(3)_c\tm SU(2)_L\tm U(1)_Y$ quantum number $\Delta_i({\bf 1,3,1})$ and $\overline\Delta_i({\bf 1,3,-1})$ and proper $Z_3$ charge assigments are embedded into the messengers,  the superpotential take the following form
\beqa
\label{superpotenial1}
W_{mess;\Delta}&\supseteq& y_S S\(\overline{\Delta}_1\Delta_1+\overline{\Delta}_2\Delta_2\)+ y_X X \overline{\Delta}_1 \Delta_2,
\eeqa

To preserve gauge coupling unification, the $\Delta_i({\bf 1,3,1})$ and $\overline\Delta_i({\bf 1,3,-1})$ messengers should be embedded into complete SU(5) representations
\beqa
{\bf 15}&=&{\bf \Delta_{S}( 6,1)_{-2/3}\oplus\Delta_T( 1,3)_{1}\oplus \Delta_{D}(3,2)_{~1/6}}~,\nn\\
\overline{\bf 15}&=&{\bf \overline{\Delta}_{S}( \bar{6},1)_{~2/3}\oplus\overline{\Delta}_T( 1,\bar{3})_{-1}\oplus\overline{\Delta}_{D}(\bar{3},{2})_{-1/6}}~.
\eeqa

So the superpotential (\ref{superpotenial1}) in terms of SU(5) representation is given by
\beqa\label{Z3NMSSM+GMSB}
W_{mess;A}&\supseteq& \sum\limits_{k=1}^{2n} y_X X\overline{\bf 15}_{a}\cdot{\bf 15}_{a}+\sum\limits_{k=1}^{n} y_S S \overline{\bf 15}_{2k-1}\cdot {\bf 15}_{2k}.
\eeqa
Although such a double-messenger-species choice of superpotential is phenomenological viable, it had been utilized in our previous model buildings, see ~\cite{Fei:1804.07335} for an example.
In this work, we choose an alternative possibility to realize NMSSM spectrum with one messenger species.

 Couplings of the form
\beqa
\label{superpotenital2}
W_{mess;\Delta}&\supseteq&  y_S S \overline{\Delta}_1\Delta_1+ y_X X \overline{\Delta}_1\Delta_1
\eeqa
 in the superpotential will trigger mixing between $X$ and $S$ via messenger loops,
generating the following Kahler potential after integrating out the messengers
\beqa
K={3y_X y_S}S X^\da\ln\(\f{X^\da X}{ M^2}\)+h.c.~,
\eeqa
which will give a tadpole term for $S$ after SUSY breaking
\beqa
{\cal L}\supseteq {3y_X y_S}(S+ S^*) M \left|\f{F_X}{M}\right|^2~.
\eeqa
Such a tadpole term can generate a suitable VEV for $\langle S\rangle$. Therefore, we adopt this possibility for GMSB.
 The superpotential (\ref{superpotenital2}) can be embedded into the following form with complete SU(5) multiplets
\beqa
\label{GNMSSM+GMSB}
W_{mess;B}&\supseteq&  y_X X\overline{\bf 15}\cdot {\bf 15}+ y_S S \overline{\bf 15}\cdot {\bf 15}.
\eeqa
 So the whole GMSB superpotential is given by
 \beqa
\label{GMSB:superpotential}
W_0&\supseteq& \f{y_{\bf 15}^u}{2}\overline{\bf 15}_{\Delta}\cdot {\bf 5}_H\cdot{\bf 5}_H+\f{y_{\bf 15}^d}{2} {\bf 15}_{\Delta}\cdot \overline{\bf 5}_H\cdot\overline{\bf 5}_H+\f{y^L_{{\bf 15};ij}}{2}{\bf 15}_{\Delta}\cdot\overline{\bf 5}_i\cdot\overline{\bf 5}_j+\la S \cdot\overline{\bf 5}_H\cdot{\bf 5}_H+\f{\kappa}{3}S^3+\cdots~\nn\\
&+&  y^u_{ij}{\bf 10}_i\cdot{\bf 10}_j\cdot{\bf 5}_H+y^d_{ij}{\bf 10}_i\cdot\overline{\bf 5}_j\cdot\overline{\bf 5}_H+W_{SB}({\bf 24},\cdots)+W_{mess;B}~,
\eeqa
with $W_{SB}({\bf 24}_H,\cdots)$ the SU(5) symmetry breaking sector, which possibly involving ${\bf 24}_H$ Higgs etc. Besides, proper doublet-triplet(D-T) splitting mechanism is assumed so that the Higgs triplets in ${\bf 5}_H$ and $\bar{\bf 5 }_H$ will be very heavy and be absent from the low energy spectrum at the messenger scale $M_{mess}$\footnote{There are many alternative model building possibilities. For example, it is possible to keep gauge coupling unification by introducing only the vector-like octet and triplet superfields. In 5D orbifold GUT model, it is possible that only the triplets zero modes can survive the orbifolding boundary conditions, which can also naturally generate D-T splitting.}. After we integrating out the messengers,  tiny Majorana neutrino masses will be generated by GNMSSM extension of type-II neutrino seesaw mechanism, the superpotential (\ref{GNMSSM+TypeII}). We should note that $Z_3$-invariant NMSSM can be generated if we adopt the superpotential (\ref{Z3NMSSM+GMSB}) instead of (\ref{GNMSSM+GMSB}).
\subsection{The analytical expressions of the soft SUSY breaking parameters at the messenger scale}
From the superpotential (\ref{GNMSSM+TypeII}), the general expressions for soft SUSY breaking parameters at the messenger scale (which is also identified to be the scale of the triplets) can be calculated with the wavefunction renormalization approach~\cite{GMSB:wavefunction}.
\bit
\item The expressions for gaugino masses
\beqa
M_{i}&=& g_i^2\f{ F_X}{2M}\f{\pa}{\pa \ln |X|}\f{1}{g_i^2}(\mu,|X|)~.
\eeqa
So we have
\beqa
M_i=-\f{ F_X}{M}\f{\al_i(\mu)}{4\pi}\Delta b_i~,\eeqa
with
\beqa
\Delta b_i=\(~7~,~7~, ~7~\)~.
\eeqa
\item The expressions for trilinear couplings
\beqa
A_0^{ijk}\equiv \f{A_{ijk}}{y_{ijk}}&=&\sum\limits_{i} \f{ F_X}{2M}\f{\pa}{\pa\ln |X|} Z({\mu}; |X|)~,\nn\\
&=&\sum\limits_{i}  \f{ F_X}{M}\f{\Delta G_i}{2}~.
\eeqa
In our convention, the anomalous dimension are expressed in the holomorphic basis~\cite{shih}
\beqa
G^i\equiv \f{d Z_{ij}}{d\ln\mu}\equiv-\f{1}{8\pi^2}\(\f{1}{2}d_{kl}^i\la^*_{ikl}\la_{jmn}Z_{km}^{-1*}Z_{ln}^{-1*}-2c_r^iZ_{ij}g_r^2\),
\eeqa
with $\Delta G\equiv G^+-G^-$ the discontinuity across the messenger threshold. Here $'G^+(G^-)'$ denote respectively the value above (below) the messenger threshold.

So we have the soft SUSY breaking trilinear couplings
\beqa
A_t&=&-\f{1}{16\pi^2}\f{ F_X}{M}2\(y^u_{\bf 15}\)^2~,\nn\\
A_b&=&-\f{1}{16\pi^2}\f{ F_X}{M}\[5\(y^L_{\bf 15;3}\)^2+2\(y^d_{\bf 15}\)^2\]~,\nn\\
A_\tau&=&-\f{1}{16\pi^2}\f{ F_X}{M}\[5\(y^L_{\bf 15;3}\)^2+2\(y^d_{\bf 15}\)^2\]~,\nn\\
A_\la&=&-\f{1}{16\pi^2}\f{ F_X}{M}\[ 15\(y_S\)^2+2\(y^u_{\bf 15}\)^2+2\(y^d_{\bf 15}\)^2\]~,\nn\\
A_\ka&=&-\f{1}{16\pi^2}\f{ F_X}{M} 45\(y_S\)^2~,\nn\\
m_{S^\pr}^2&=&-\f{\mu^\pr}{16\pi^2}\f{ F_X}{M} 30\(y_S\)^2~,\nn\\
\xi_S&=&-\f{\xi_F}{16\pi^2}\f{ F_X}{M} 15\(y_S\)^2~,\nn\\
m_3^2&=&-\f{\mu}{16\pi^2}\f{F_X}{M} \[2\(y^u_{\bf 15}\)^2+2\(y^d_{\bf 15}\)^2\]~,
\eeqa
Here we neglect possible RGE effects of $y_{H_u H_u \overline{\Delta}}$ etc between the GUT scale and the messenger scale.

\item The soft SUSY masses are given as
\beqa
m^2_{soft}&=&-\f{F^2_X}{4 M^2}\f{\pa}{\pa(\ln |X|)^2} \ln \[Z_i(\mu,X,T)\],\nn\\
&=&-\f{F^2_X}{4 M^2}\[\f{\pa}{\pa \ln M}\Delta G-\f{\pa}{\pa \ln M}G^-(M,\ln M)\].
\eeqa

The expressions for soft scalar masses are rather lengthy. So we collect their expressions in appendix \ref{appendix:A}.
\eit

As discussed in section \ref{type II}, the soft SUSY breaking trilinear term can give subleading contribution to Majorana neutrino mass via type-II seesaw mechanism. We require the knowledge of trilinear scalar coupling $\tl{\Delta}_T-H_d-H_d$ in GMSB. However, there are no contributions to trilinear couplings $\tl{\Delta}_T-H_d-H_d$ at the $M_{Mess}$ scale in GMSB. So the trilinear soft term contribution to type-II seeesaw neutrino masses will not play a role.

\subsection{Numerical constraints on type-II neutrino seesaw mechanism extension of NMSSM from GMSB}

Lacking gauge interactions for $S$, ordinary GMSB predicts vanishing trilinear couplings $A_\ka, A_\la$ and vanishing $m_S^2$. Therefore, it can not predict realistic low energy NMSSM spectrum unless additional Yukawa mediation contributions are present~\cite{NMSSM:GMSB}. Fortunately, because of the new interactions involving $H_u,H_d,S$ and triplets, additional Yukawa deflection contributions related to type-II neutrino seesaw can lead to new contributions to trilinear couplings and soft scalar masses. Therefore, phenomenological viable parameters can be possible in our scenario.

 In ordinary setting, the spurion $X$ is normalized so that $y_X=1$. Due to possible mixing between $X$ and $S$ through messengers in ${\bf 15}$ representation of SU(5), tadpole terms in the scalar potential of $S$ can be generated as
 \beqa
 \xi_S=15 y_X y_S\f{F_X^2}{M}~.
 \eeqa
 So general NMSSM soft SUSY breaking parameters will appear in the GMSB scenario. Besides, we set  $\xi_F=\mu^\pr=0$ to keep the predictive power of the scenario.

 The free parameters in this scenario are given as
 \beqa
 \f{F_X}{M_{mess}},~~~M_{mess},~~y^L_{\bf 15;a},~~y^d_{\bf 15},~~~y^u_{\bf 15},~~~y_S,~~~\la,~~~\ka.
 \eeqa

For simplicity, we adopt the universal inputs for the new Yukawa couplings at the messenger scale
\beqas
y^L_{\bf 15;a}=\la_0~,~y^d_{\bf 15}=y^u_{\bf 15}=\la_1~,~y_S=\la_2~.
\eeqas

The soft SUSY masses $m_{H_u}^2,m_{H_d}^2,m_S^2$ can be reformulated into $\mu,\tan\beta,M_Z^2$ by the minimum conditions of the scalar potential
  \beqa
  M_A^2=\f{2\mu_{eff}}{\sin2\beta}B_{eff}~, ~~\mu_{eff}\equiv\la\langle s\rangle~,~~~B_{eff}=(A_\la+\ka \langle s\rangle).
  \eeqa
  In our numerical study, $\kappa$ is a free parameter while $\tan\beta$ is not. This choice is different to ordinary numerical setting in NMSSM in which $\tan\beta$ is free while $\kappa$ is a derived quantity~\cite{ellwanger:0612134}. Such a choice can be convenient for those predictable NMSSM models from top-down approach. A guess of $\tan\beta$ is made to obtain the relevant Yukawa couplings $y_t,y_b$ at the EW scale. After RGE evolving up to the messenger scale, the whole soft SUSY breaking parameters at the messenger scale can be determined. Low energy $\tan\beta$ can be obtained iteratively with such a spectrum from the minimization conditions of the Higgs potential. Obtaining an iteratively stable $\tan\beta$ indicates that the EWSB conditions are satisfied by the model input.

It can be calculated that $2n$ generations of $\overline{\bf 15},{\bf 15}$ superfields of SU(5) will contribute $\Delta b_i=14n$ to the gauge beta functions. Perturbativity of the gauge coupling at the unification scale requires the combination~\cite{1811.12336}
\beqa
\label{perturb:bound}
\delta=-\f{14n}{2\pi}\ln \f{M_{GUT}}{M_{mess}}~,
\eeqa
to satisfy
\beqa
|\delta|\lesssim 24.3~,
\eeqa
with $M_{GUT}$ and $M_{mess}$ the GUT scale and messenger scale, respectively.
So the messenger scale need to satisfy $M_{mess}\gtrsim 10^{11}$ GeV for $n=1$ and $M_{mess} \gtrsim 10^{14}$ GeV for $n=2$.

    We use NMSSMTools 5.5.0~\cite{NMSSMTools} to scan the whole parameter space. Randomly scan in combine with MCMC method are used. We interest in relatively large values of $\la$ in order to increase the tree-level mass of the 125 GeV CP-even Higgs boson. Besides, the couplings $\la_0,\la_1,\la_2$ should be pertubative and $\la,\ka$ should satisfy the perturbative bound $\la^2+\ka^2\lesssim 0.7$.
     The parameters are chosen to lie the following range
 \beqa
&&~~ 10^{11}~ {\rm GeV}< M_{mess}\equiv m_T<2.0\tm 10^{14} ~{\rm GeV}~,~~ 10~ {\rm TeV} < \f{F_X}{M}< 500~ {\rm TeV}~,~\nn\\
&&~~~~~~~~~~~0<\la_0,\la_1,\la_2<\sqrt{4\pi}~,~~~~~~~~~~~~~~~~~0.1<\la,\ka <0.7~.
 \eeqa
 The coupling $\la_1$, which is just the $y_\Delta^u$ in equation (\ref{GNMSSM+TypeII}), should not be too small. Otherwise, very large $(m_T/y_\Delta^u)$ factor will lead to large $\(m^2_{\tl{L}}\)_{ij}$, which may exceed the current bounds on $Br(l_i\ra l_j\ga)$ even if the leading-log contributions are not enhanced by the log factor. Conservative bound $(m_T/y_\Delta^u)\lesssim 0.6\tm 10^{14}$ GeV, obtained numerically in ~\cite{typeII+GMSB}, can be imposed in subsequent numerical results to ensure that our scenarios can be safely compatible with $\mu\ra e\ga$ constraints etc. Therefore, we have an upper bound $m_T\equiv M_{mess}\lesssim 2.0 \tm 10^{14}{\rm  GeV}$ for $y_\Delta^u\sim \sqrt{4\pi}$. This upper bound of $M_{mess}$ also safely lie below the GUT scale. So we choose the conservative upper bound of $M_{mess}$ to be $2.0\tm 10^{14} {\rm GeV}$ in our numerical scan. The lower bound of $M_{mess}$ comes from the perturbative requirements of gauge couplings below $M_{GUT}$, which can be seen from the discussion below eqn(\ref{perturb:bound}).

In addition to the constraints from neutrino masses
 \beqa
 \label{neutrino:part}
\left|\(m_\nu\)_{ij}\right|\approx y^L_{ij}\[ y_\Delta^u \f{v_u^2}{m_T}\] \lesssim ~0.1~ {\rm eV},
\eeqa
 we also impose the following constraints in our numerical scan
\bit
\item (I) The conservative lower bounds from current LHC constraints on SUSY particles~\cite{atlas:gluino,atlas:stop}:
    \bit
    \item  Light stop mass: $m_{\tl{t}_1} \gtrsim 0.85$ TeV.
    \item  Gluino mass: $m_{\tl{g}} \gtrsim 1.5\sim 1.9 $ TeV.
    \item  Light sbottom mass $m_{\tl{b}_1} \gtrsim 0.84$ TeV.
    \item  Degenerated first two generation squarks $m_{\tl{q}} \gtrsim 1.0 \sim 1.4$ TeV.
    \eit
\item (II) We impose the following lower bounds for neutralinos and charginos, including the invisible decay bounds for $Z$-boson. The most stringent constraints of LEP~\cite{LEP} require $m_{\tl{\chi}^\pm}> 103.5 {\rm GeV}$ and the invisible decay width $\Gamma(Z\ra \tl{\chi}_0\tl{\chi}_0)<1.71~{\rm MeV}$,
     which is consistent with the $2\sigma$ precision EW measurement constraints $\Gamma^{non-SM}_{inv}< 2.0~{\rm MeV}$.

\item  (III) Recent flavor constraints from rare B meson decays ~\cite{B-physics}:
  \beqa
  && 0.85\tm 10^{-4} < Br(B^+\ra \tau^+\nu) < 2.89\tm 10^{-4}~,\nn\\
  &&1.7\tm 10^{-9} < Br(B_s\ra \mu^+ \mu^-) < 4.5\tm 10^{-9}~,\nn\\
  && 2.99\tm 10^{-4} < Br(B_S\ra X_s \gamma) < 3.87\tm 10^{-4}~.
   \eeqa

\item (IV) The CP-even component $S_2$ in the Goldstone-$'eaten'$ combination of $H_u$ and $H_d$ doublets corresponds to the SM Higgs boson. The $S_2$ dominated CP-even Higgs should lie in the combined mass range for the Higgs boson: $125\pm  3 {\rm GeV}$ from ATLAS and CMS data, where the width of the band is given by the theoretical uncertainty of the Higgs mass calculation. The uncertainty is 3 GeV instead of default 2 GeV because large $\lambda$ may induce additional ${\cal O}(1)$ GeV correction to $m_h$ at two-loop level~\cite{NMSSM:higgs2loop}, which is not included in the NMSSMTools.

\eit

      It is known that gravitino $\tl{G}$ will be much lighter in GMSB than that in mSUGRA and in general will be the lightest supersymmetric particle(LSP). Such a light gravitino is also motivated by cosmology since it can evade the gravitino problem.  The interaction of goldstino component of gravitino is $1/F_X$ instead of $1/M_{Pl}$. If gravitinos are in thermal equilibrium at early times and freeze out at the temperature $T_f$ , their relic density is~\cite{gravitino:relic}
    \beqa
\Omega_{\tl{G}}h^2=\f{m_{\tl{G}}}{{\rm keV}}\f{100}{g_*(T_f)}.
    \eeqa
   In order to obtain the required dark matter(DM) relic density, one needs to adjust the reheating temperature as a function of the gravitino mass. Besides, it is shown in ~\cite{gravitino:reheating} that the late decay of the lightest messenger to visible sector particles can induce a substantial amount of entropy production which would result in the dilution of the predicted gravitino abundance. As a result, one would obtain suitable gravitino dark matter for arbitrarily high reheating temperatures. Due to the flexibility of the theory, we do not impose the DM relic density constraints in our GMSB scenario.

\begin{figure}
\begin{center}
\includegraphics[width=2.8in]{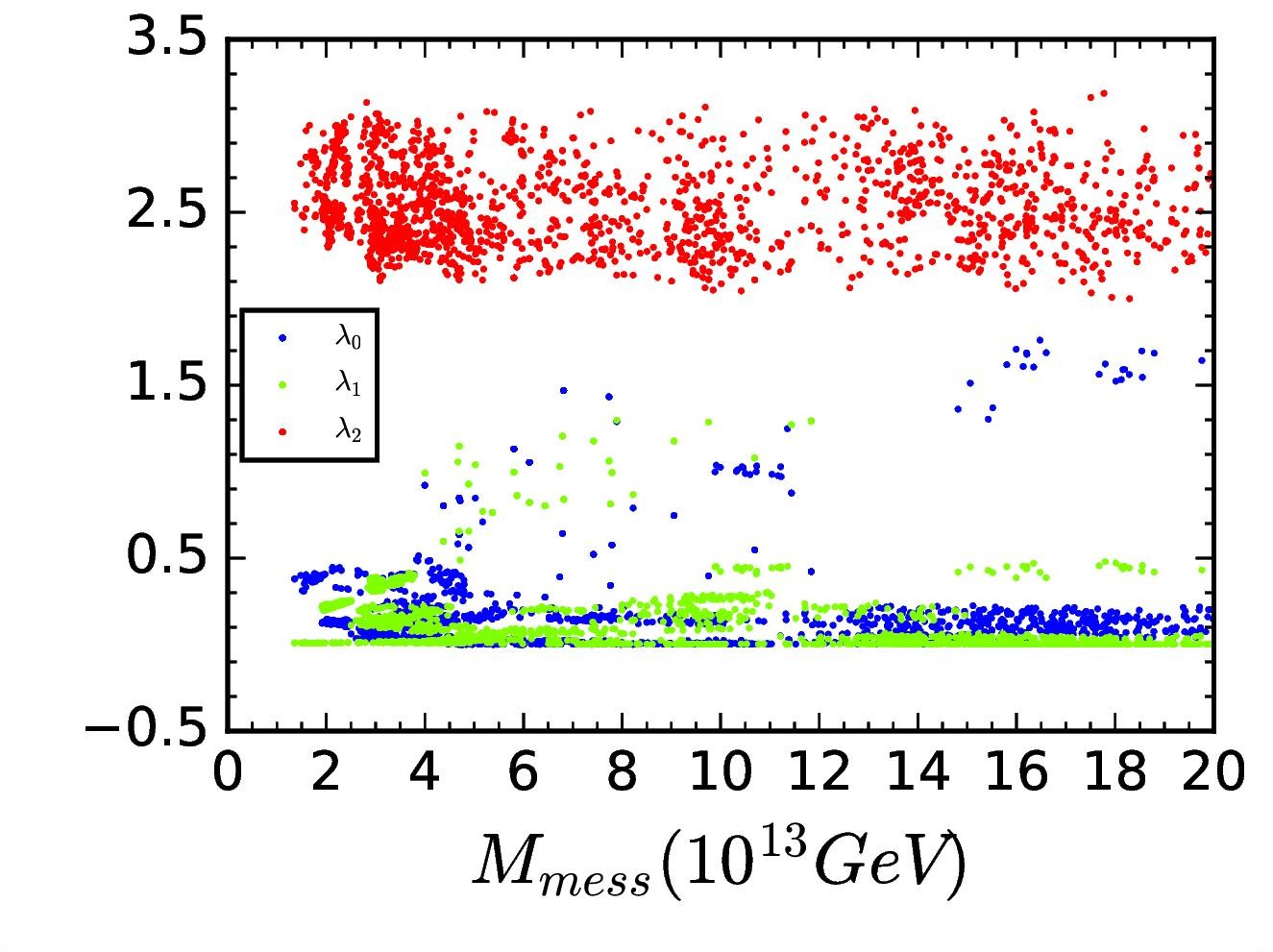}
\includegraphics[width=2.8in]{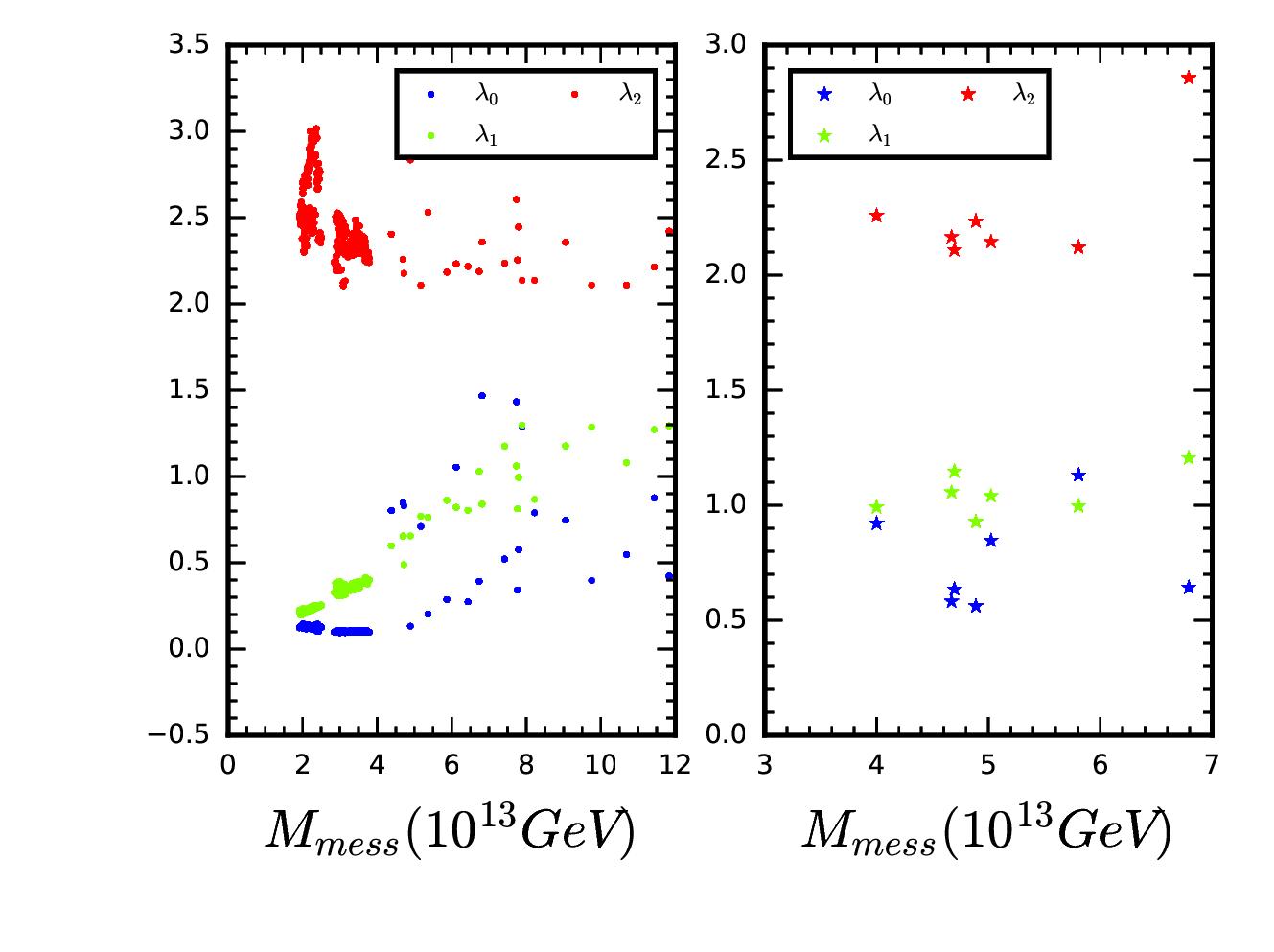}\\
\includegraphics[width=2.8in]{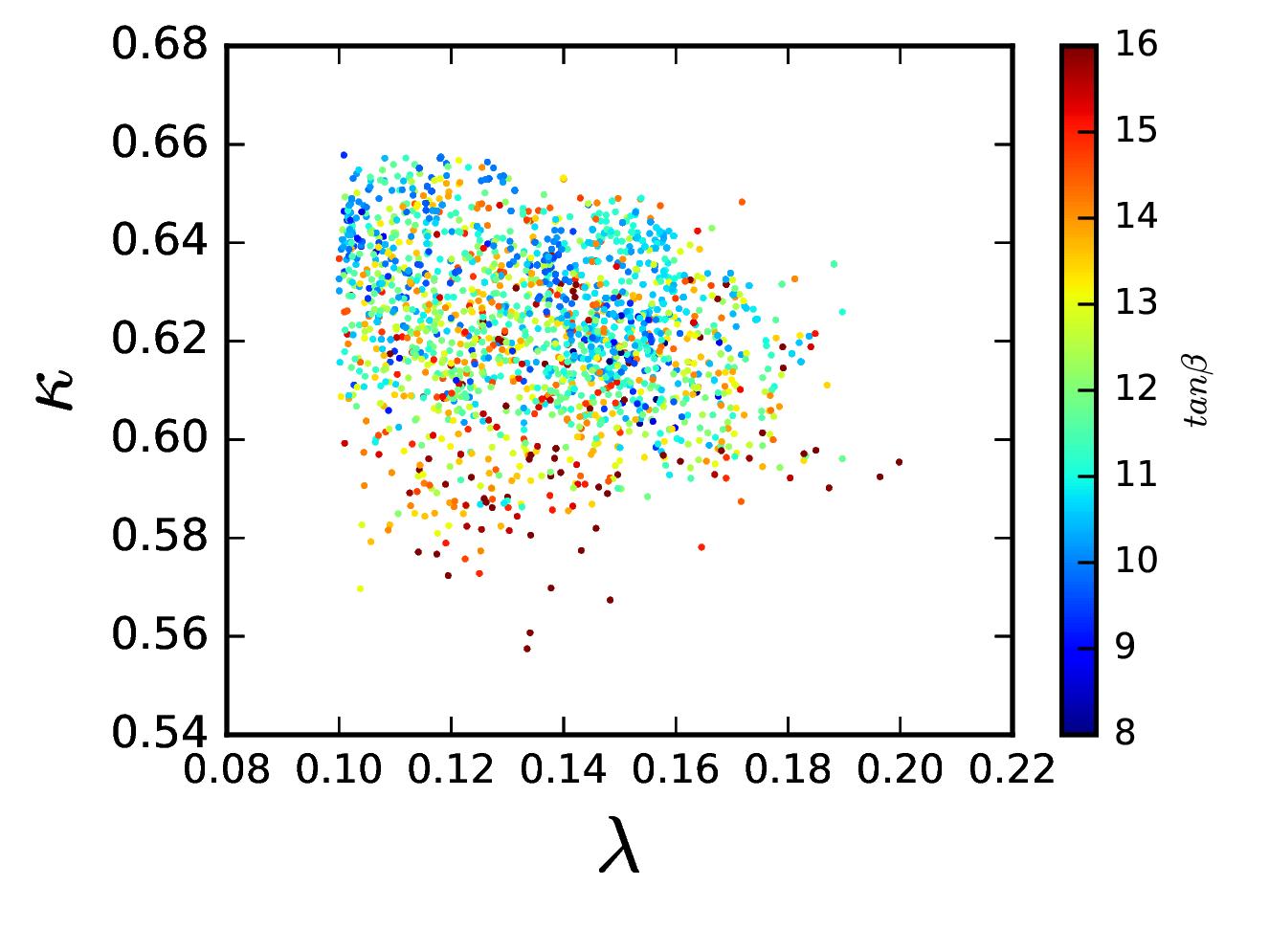}
\includegraphics[width=2.8in]{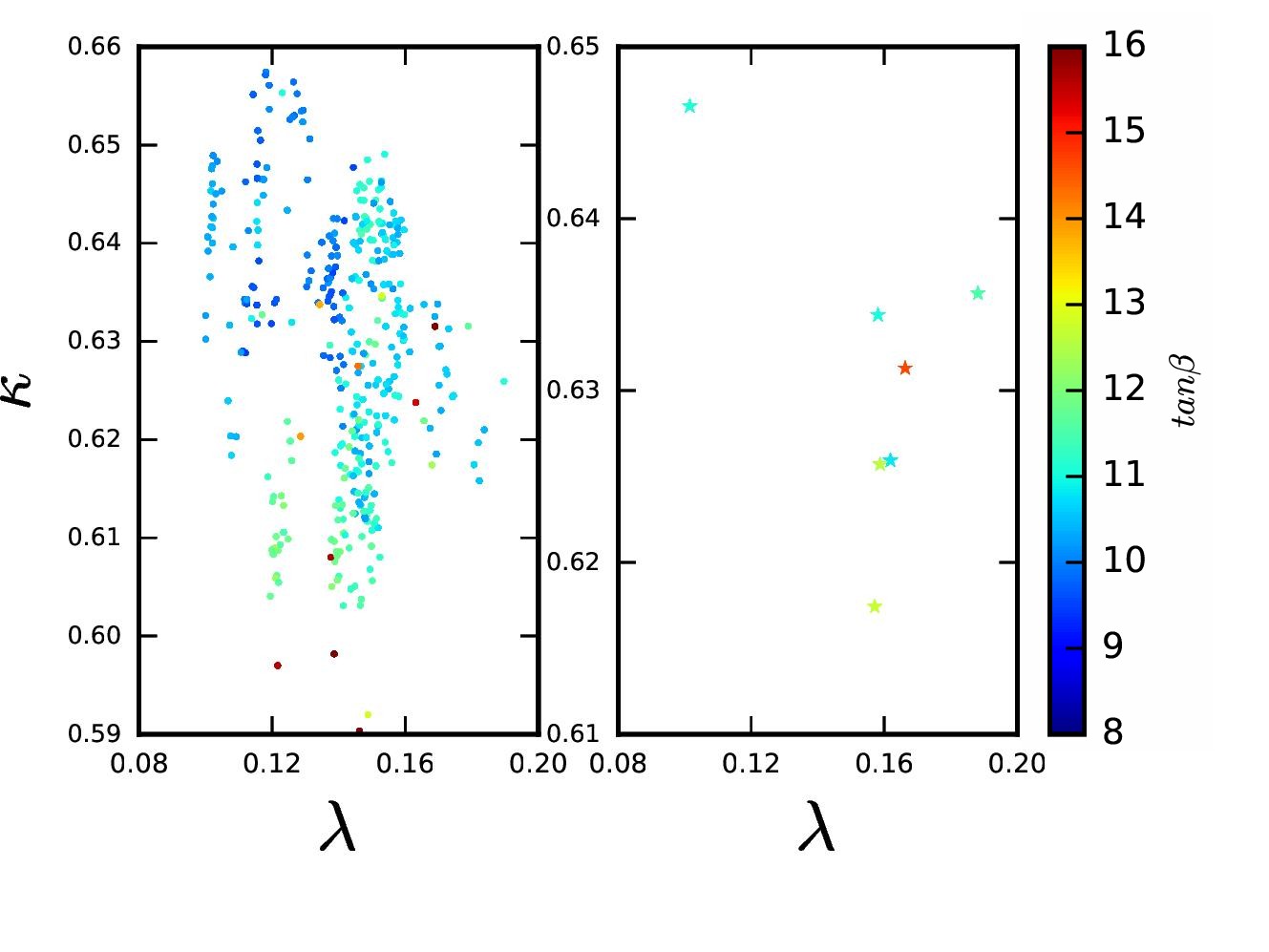}\\
\includegraphics[width=2.8in]{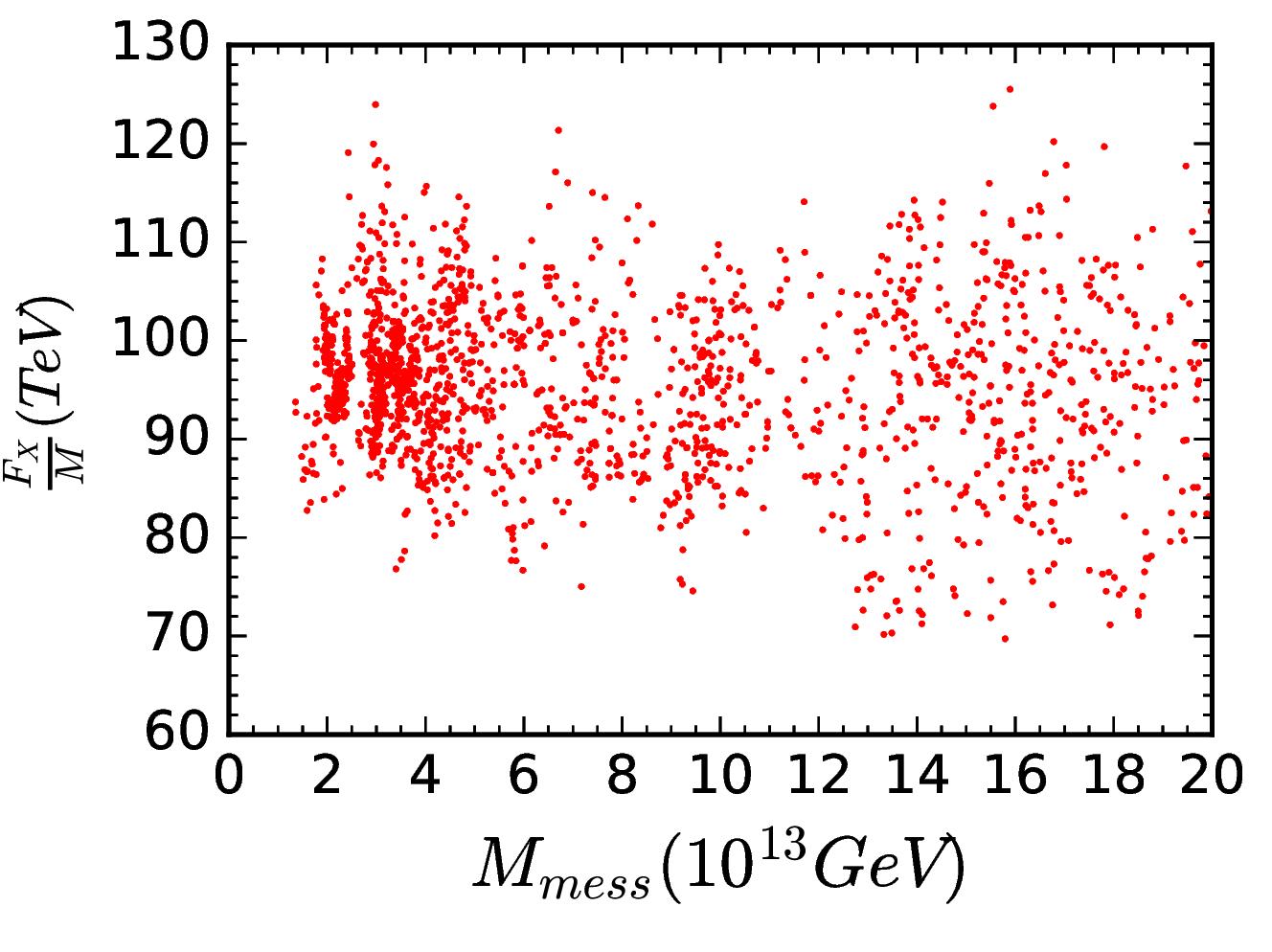}
\includegraphics[width=2.8in]{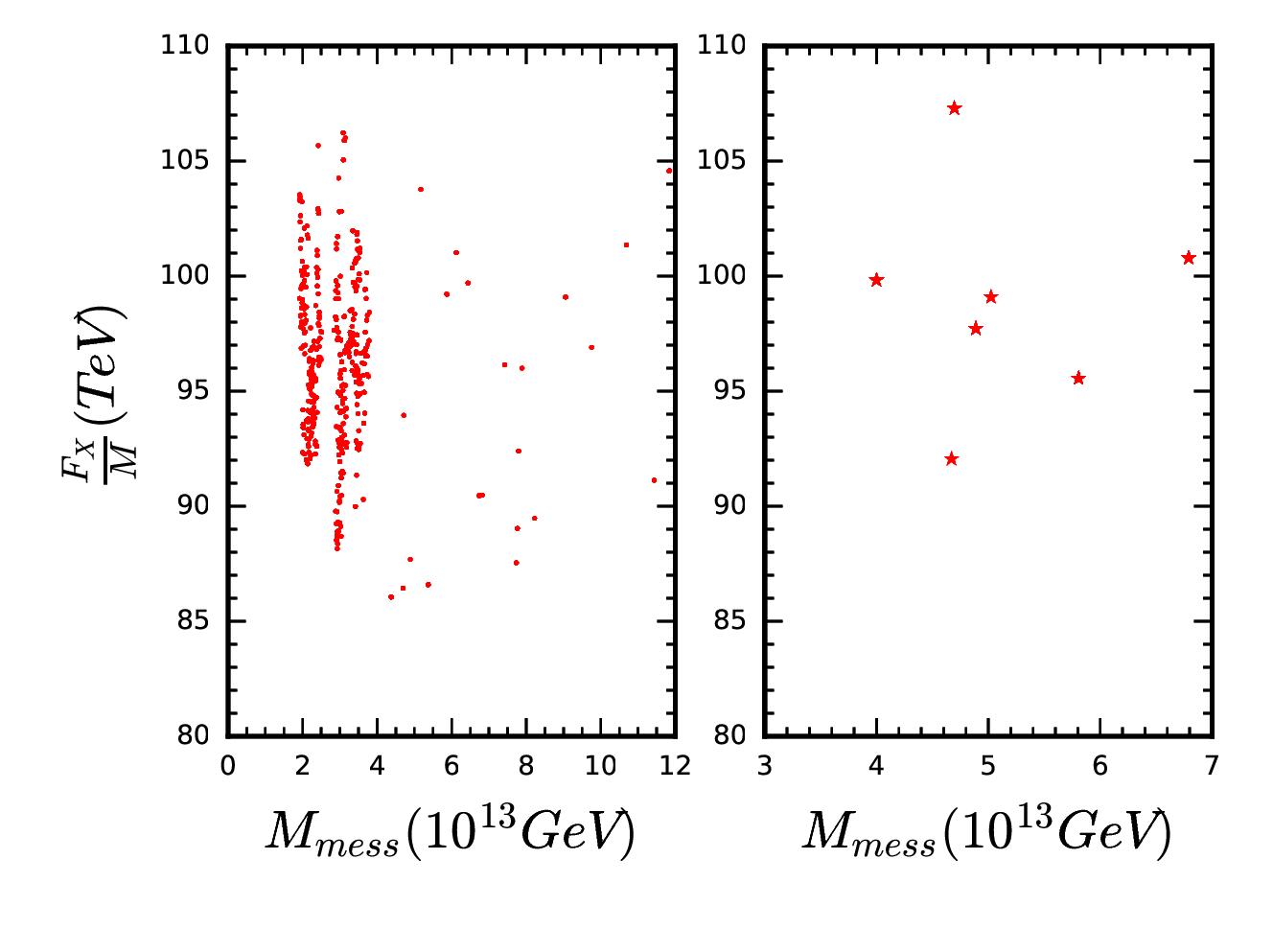}
\includegraphics[width=2.8in]{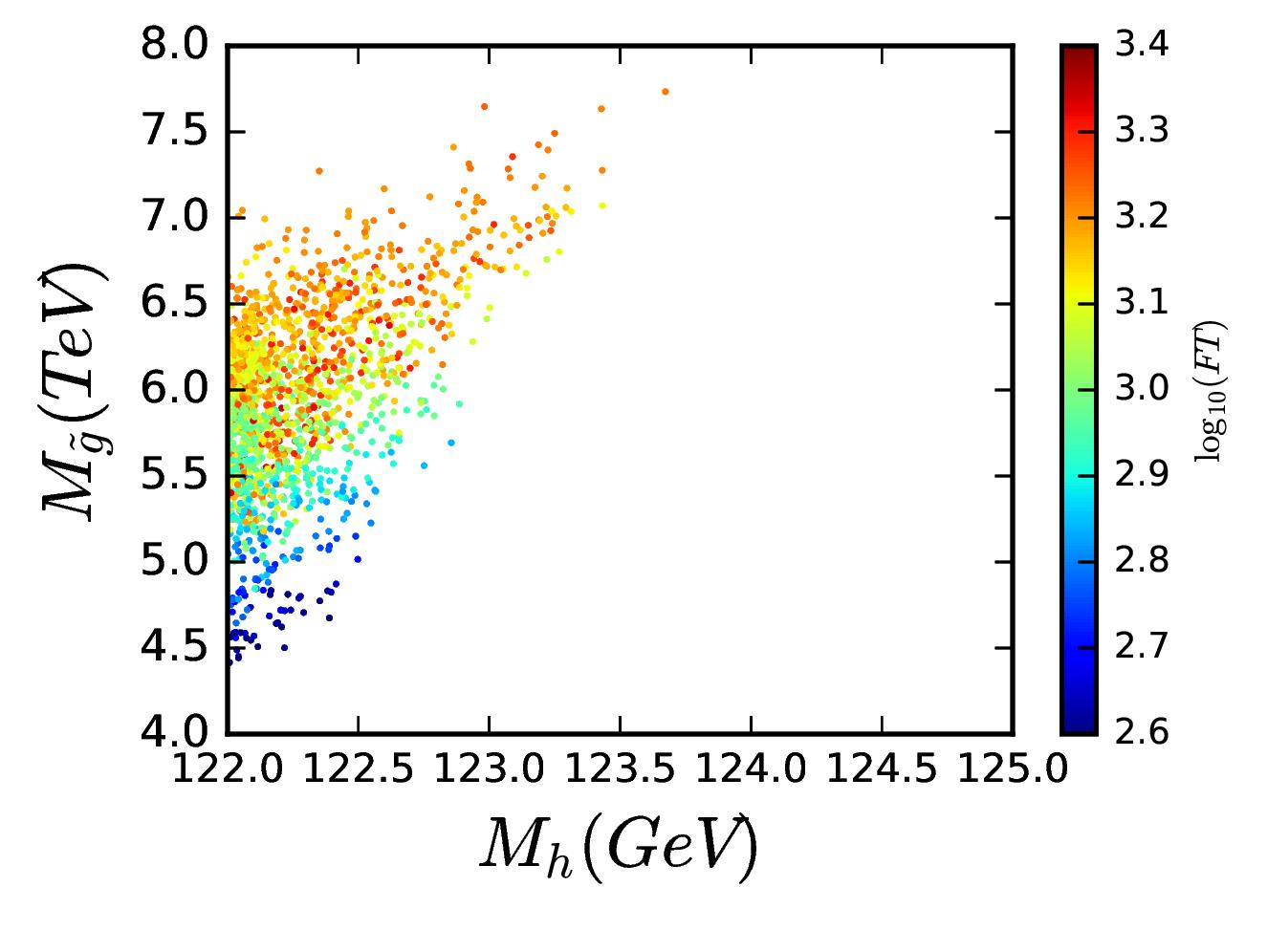}
\includegraphics[width=2.8in]{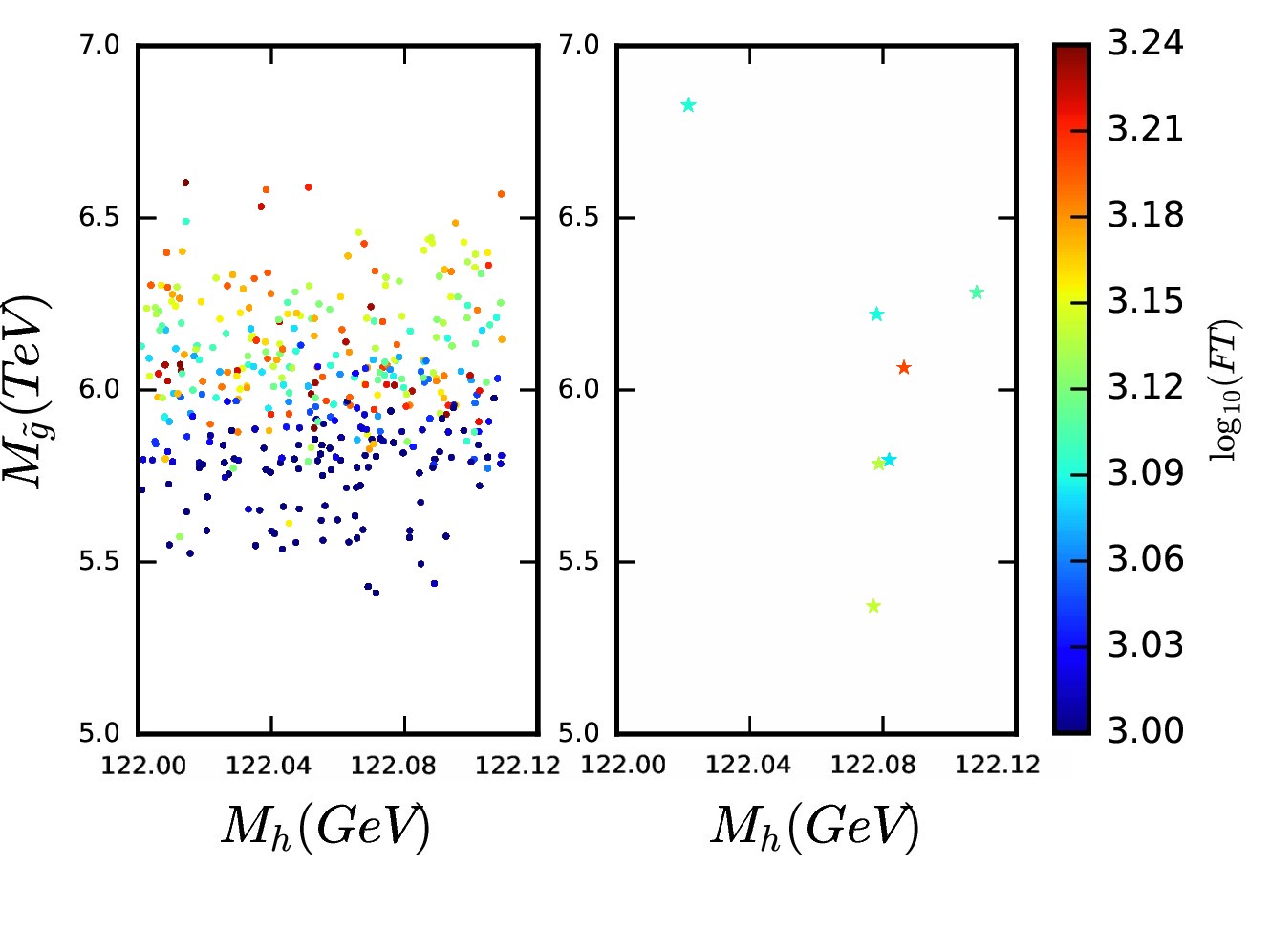}\\
\end{center}
\vspace{-.5cm}
\caption{Survived points that can satisfy the EWSB conditions and the constraints from (I) to (V) in type-II neutrino seesaw mechanism extension of NMSSM from GMSB. The 125 GeV Higgs is found to be the lightest CP-even scalar for all the survived points. The BG fine tuning measures are also shown in different colors. In the right panels, we show the survived points with additional LFV bounds $(m_T/y_\Delta^u)\lesssim 1.0\tm 10^{14}$ GeV (left) and $0.6\tm 10^{14}$ GeV(right), respectively.}
\label{fig5}
\end{figure}

 To illustrate the constraints from LFV processes $l_i\ra l_j \ga$, we show in the right panels of fig.\ref{fig5} the survived points with additional LFV bounds $(m_T/y_\Delta^u)\lesssim 1.0\tm 10^{14}$ GeV (left) and $0.6\tm 10^{14}$ GeV(right), respectively.  We have the following discussions related to our numerical results
\bit
\item  It is fairly nontrivial to check if successful EWSB condition is indeed fulfilled.
The survived points after imposing the EWSB constraints and the bounds from (I) to (IV) are shown in fig.\ref{fig5}.
 As shown in upper left panel of fig.\ref{fig5}, numerical results indicate that the non-trivial couplings $\la_0,\la_1,\la_2$, especially $\la_2\in[1.8,3.3]$, are required to obtain realistic low energy NMSSM spectrum. Non-vanishing $\la_2$, which determines the couplings between $S$ and the messengers, is necessary to give sizable contributions to the trilinear couplings $A_\ka$ and $m_S^2$, which receive no additional contributions from pure GMSB. Such Yukawa mediation contributions, whose sizes need to be of order the EW scale, are indispensable to satisfy the EWSB conditions and could determine the size of $\la_2$ to be of ${\cal O}(1)$. The couplings $\lambda_2$ can also contribute to the Higgs masses. Constraints from the neutrino masses on $\la_0,\la_1$ and $M_{mess}$ are fairly mild because the combination eqn(\ref{neutrino:part}) can easily be satisfied in the allowed range of the parameters. It can be seen that the scale of the triplet in GMSB scenario are constrained to lie above $10^{13}$ GeV. We checked that lower value of $M_{mess}$ can not survive the bounds from Higgs mass and LHC data. No additional upper bounds for $M_{mess}$ (other than $2.0\tm 10^{14}{\rm GeV}$) are found from constraints (I) to (IV). From the upper right panels of fig.\ref{fig5}, we can see that LFV bounds can be fairly restrictive. Many otherwise survived points are ruled out by the $m_T/y_\Delta^u$ bound. If we choose $(m_T/y_\Delta^u)\lesssim 0.6\tm 10^{14}$, an upper bound for the messenger scale $M_{mess}\lesssim 6.9\tm 10^{13} $ GeV can be obtained from our numerical results.

\item  Without the constraints on $m_T/y_\Delta^u$, the values of $\ka$ should lie between 0.54 to 0.66 (see the left panel in the second row of fig.\ref{fig5}). It is also clear that the allowed ranges of $\lambda$ and the iteratively obtained (from EWSB conditions) $\tan\beta$, are found to lie between [0.1,0.2] and [8,16], respectively. The value of $F_X/M$ determines the whole scale of the soft SUSY breaking spectrum, including the top squark masses and the scale of the trilinear coupling $A_t$. From the left panel in the third row of fig.\ref{fig5}, we can see that $F_X/M$ should take the values between $70$ TeV to $130$ TeV to generate sparticles masses of order $1\sim 10$ TeV.

Again, it is clear from the middle right panels that the bounds from $m_T/y_\Delta^u$ can impose stringent constraints on the otherwise survived parameters. If we choose $(m_T/y_\Delta^u)\lesssim 0.6\tm 10^{14}$, the values of $\ka$ are constrained to lie between $0.61$ and $0.65$ while the values of $F_X/M$ should lie between $90$ TeV to $110$ TeV.

\item From the lower left panel of fig.\ref{fig5}, the gluinos are constrained to lie between 4.5 TeV to 8 TeV, which can be accessible only in the future VLHC with $\sqrt{s}=100$ TeV. It is also clear that our scenario can successfully account for the 125 GeV Higgs boson in the case that the 125 GeV Higgs is the lightest CP-even scalar.The Higgs mass in NMSSM can be approximately given by~\cite{NMSSM}
  \beqa
m^2_{h} &\simeq& M_Z^2 \cos^2 2\beta
+\la^2 v^2 \sin^2 2\beta -\frac{\la^2}{\ka^2}v^2(\la-\ka\sin2\beta)^2\nn \\
&+& \frac{3m_t^4}{4\pi^2 v^2}
\left[\ln\left(\frac{m_{\tl{T}}^2}{m_t^2}\right) +\frac{A_t^2}{m_{\tl{T}}^2}
\left(1-\frac{A_t^2}{12m_{\tl{T}}^2}\right)\right]~,
\label{higgs:NMSSM}
\eeqa
 with $v \approx 174 ~{\rm GeV}$, $m_{\tl{T}}^2=m_{U_3}^2$ and $A_t$ the stop trilinear coupling.
 As the survived points require large $\ka$, small $\la$ and intermediate $\tan\beta$, the NMSSM specific tree-level contribution $\la^2 v^2\sin^2 2\beta$ to Higgs mass is always small. Besides, the mixing with the singlet scalar will provide destructive contributions to Higgs mass, which can be seen in eqn.(\ref{higgs:NMSSM}). Therefore, large $A_t$ or heavy stop masses are still needed in this scenario to accommodate 125 GeV Higgs. Fortunately, due to the new contributions to $A_t$ from type-II neutrino seesaw specific interactions, the 125 GeV Higgs can be successfully obtained by some portion of input parameters. The survived ranges of $F_X/M$ can just lead to such TeV scale stops and $A_t$ term. We also note that the value $A_t/m_{\tl{T}}$ lies typically away from the maximal mixing value $A_t/m_{\tl{T}}\simeq \pm \sqrt{6}$. So, the contribution from the second term of second line in eqn.(\ref{higgs:NMSSM}) is very small, necessitating large contributions from the $\ln(m_{\tl{T}}/m_t)$ term with relatively heavy stops.
Although the 125 GeV Higgs boson can be either the lightest or the next-to-lightest CP-even scalar, our numerical results indicate that it can only be the lightest CP-even scalar in this scenario.
A benchmark point is given in Table \ref{Delta} to illustrate the soft spectrum of our economical type-II neutrino seesaw mechanism extension of NMSSM from GMSB.
\begin{table}[htbp]
 \caption{Benchmark point for our economical type-II neutrino seesaw mechanism extension of NMSSM from GMSB. All mass parameters are in the unit of GeV.}

 \begin{tabular}{|c|c|c|c|c|c|}
 \hline
 \hline
 ${F_X}/{M_{mess}}$&$89968.746$&$M_{mess}$&$0.317\tm10^{14}$&$\la_0$&$0.101$\\
 \hline
$\la_1$&$0.216$&$\la_2$&$2.718$&$\la$&$0.111$\\
 \hline
 $\ka$&$0.623$&&&&\\
 \hline
 \hline
 $\tan{\beta}$&$12.693$&$A_\la$&$-14876.749$&$A_\ka$&$-42738.328$\\
 \hline
 $A_t$&$4368.885$&$A_b$&$6661.326$&$A_\tau$&$974.974$\\
 \hline
 $m_{h_1}$&$122.148$&$m_{h_2}$&$2307.837$&$m_{h_3}$&$14038.417$\\
 \hline
 $m_{a_1}$&$2307.809$&$m_{a_2}$&$45667.152$&$m_h^{\pm}$&$2309.024$\\
 \hline
 $m_{\tl{d}_{L}}$&$5404.684$&$m_{\tl{d}_{R}}$&$5210.315$&$m_{\tl{u}_{L}}$&$5404.147$\\
 \hline
 $m_{\tl{u}_{R}}$&$5239.518$&$m_{\tl{s}_{L}}$&$5404.684$&$m_{\tl{s}_{R}}$&$5210.315$\\
 \hline
$m_{\tl{c}_{L}}$&$5404.147$&$m_{\tl{c}_{R}}$&$5239.518$&$m_{\tl{b}_{1}}$&$4809.792$\\
 \hline
$m_{\tl{b}_{2}}$&$5069.749$&$m_{\tl{t}_{1}}$&$4078.332$&$m_{\tl{t}_{2}}$&$4815.497$\\
 \hline
$m_{\tl{e}_{L}}$&$1584.089$&$m_{\tl{e}_{R}}$&$881.781$&$m_{\tl{\nu_e}}$&$1582.271$\\
 \hline
$m_{\tl{\mu}_{L}}$&$1584.089$&$m_{\tl{\mu}_{R}}$&$881.781$&$m_{\tl{\nu_{\mu}}}$&$1582.271$\\
 \hline
$m_{\tl{\tau}_{1}}$&$836.131$&$m_{\tl{\tau}_{2}}$&$1572.890$&$m_{\tl{\nu_{\tau}}}$&$1570.408$\\
 \hline
$m_{\tl{\chi}_{1}^{0}}$&$-831.180$&$m_{\tl{\chi}_{2}^{0}}$&$-1672.308$&$m_{\tl{\chi}_{3}^{0}}$&$2766.952$\\
 \hline
 $m_{\tl{\chi}_{4}^{0}}$&$-2768.720$&$m_{\tl{\chi}_{5}^{0}}$&$31068.790$&$m_{\tl{\chi}_{1}^{\pm}}$&$-1672.308$\\
 \hline
 $m_{\tl{\chi}_{2}^{\pm}}$&$2769.152$&$\mu_{eff}$&$2728.527$&$m_{\tl{g}}$&$5634.875$\\
 \hline
 \hline
 \end{tabular}
\label{Delta}
\end{table}

\item  The Barbieri-Giudice fine-tuning(FT) measure with respect to certain input parameter $'a'$ is defined as~\cite{BGFT}
          \beqa
          \Delta_a\equiv \left|\f{\pa \ln M_Z^2}{\pa \ln a}\right|~,
          \eeqa
 while the total fine-tuning is defined to be $ \Delta= \max\limits_{a}(\Delta_a)$ with $\{a\}$ the set of parameters defined at the input scale.

 The Barbieri-Giudice FT measures of our scenario are shown in the the lower panels. Without the constraints on $m_T/y_\Delta^u$, the BGFT satisfies $100 \lesssim\Delta\lesssim 1000$ . Especially, in the most interesting region where $m_h\gtrsim 123 {\rm GeV}$, the BGFT are of order $1000$. The BGFT can be as low as 100 in low gluino mass regions. As the gluino mass is determined by $F_X/M$, which set all the soft SUSY mass scale, lighter $m_{\tl{g}}$ in general indicates lighter stop, reducing the FT involved. The survived points with LFV bounds are shown in the lower right panels of fig.\ref{fig5}. The predicted Higgs mass can not exceed $122.2$ GeV in such cases. If we choose $(m_T/y_\Delta^u)\lesssim 0.6\tm 10^{14}$, the predicted Higgs mass should lie near 122 GeV with the BGFT of order 1000.

 We should note BGFT in general will overestimate the fine-tuning~\cite{FT:measure}.  In fact, even if the low energy effective theory looks fine-tuned, the high scale correlations present in the ultimate theory lead to little or no fine-tuning.

 \item  Although it is possible for lightest stau to be the next-to-lightest supersymmetric particle(NLSP) in ordinary GMSB, we checked that the NLSP in our GMSB scenario will always be the lightest neutralino. Its dominant decay mode is $\tl{\chi}_1^0\ra \ga+\tl{G}$. We know that the triplet messenger scale should be very high (of order $10^{14}$ GeV) to accommodate the type-II neutrino seesaw mechanism. Therefore, to obtain TeV scale SUSY particle, the SUSY breaking $\sqrt{F_X}$ should be of order $10^{8}\sim 10^9$ GeV. As the parameter $1/F_X$ determines the lifetime of the NLSP decaying into gravitino, the average distance traveled by neutralino NLSP can be large so as that it decays outside the detector and therefore behaves like a stable particle. So the collider signatures closely resemble those of the ordinary supersymmetric scenarios with a stable neutralino.

\eit

\section{\label{AMSB}Type-II neutrino seesaw mechanism extension of NMSSM from AMSB}

To generate realistic EWSB in NMSSM, soft SUSY breaking parameters relating to singlet $S$ are necessarily present. As the gauge singlet receives no contributions from pure gauge mediation, additional Yukawa mediation contributions should be present in addition to pure GMSB contributions, complicating the relevant model building. Besides,  the numerical results in previous section indicate that it is still hard to interpret the 125 GeV Higgs mass even in (G)NMSSM because of the small NMSSM-specific tree-level contributions to Higgs mass constrained from EWSB conditions.

To simplify the previous problems in GMSB, we can move to the predictive AMSB scenario, in which AMSB contributions to $m^2_S,A_\ka,A_\la$ are naturally present. Unfortunately, the minimal AMSB scenario predicts negative slepton square masses and must be extended. The most elegant solution to tachyonic slepton is the deflected  AMSB~\cite{dAMSB} scenario in which additional messenger sectors are introduced to deflect the AMSB trajectory and lead to positive slepton mass by additional gauge or Yukawa mediation contribution. The triplets, which are required in type-II neutrino seesaw mechanism, can naturally be fitted into the messenger sector.
\subsection{Theoretical setting of our model and the soft SUSY breaking parameters}
In AMSB, the GMSB contributions of the messengers will cancel the change of AMSB contributions if simple mass thresholds for messengers are present. To evade such a difficulty, a pseudo-moduli field $X$ can be introduced with its VEV $\langle X\rangle=M+\theta^2 F_X$ to determine the messenger threshold as well as the SUSY breaking order parameter. A deflection parameter $'d'$ , which characterizes the deviation from the ordinary AMSB trajectory, can be introduced and its concrete value will depend on the form of the pseudo-moduli superpotential $W(X)$. Positive slepton masses can be achieved with either sign of deflection parameter $'d'$. The superpotential in this scenario can also be written as the form in eqn(\ref{GMSB:superpotential}). The $W_{mess;B}$ is replaced by
\beqa
W_2=y_X X  \overline{\bf 15}\cdot{\bf 15}+W(X)~,
\eeqa
within which the coupling between $S$ and $\overline{\bf 15},{\bf 15}$ is absent, simplifying the AMSB model building. So $Z_3$-invariant NMSSM can be adopted here without the needs of double messenger species. Expression of $W(X)$ can be fairly generic and leads to a deflection parameter of either sign given by
\beqa
d \equiv \f{F_X}{M F_\phi}-1~.
\eeqa
After integrating out the messengers, the type-II neutrino seesaw mechanism extension of NMSSM can be obtained with the NMSSM superpotential taking the $Z_3$ invariant form. We can calculate the soft SUSY breaking parameters following the approach in our previous works~\cite{MM:DAMSB}.
\bit
\item  The soft gaugino mass is given at the messenger scale by
\beqa
M_{i}(M_{mess})&=& g_i^2\(\f{F_\phi}{2}\f{\pa}{\pa \ln\mu}-\f{d F_\phi}{2}\f{\pa}{\pa \ln |X|}\)\f{1}{g_i^2}(\mu,|X|,T)~.
\eeqa
So the gaugino masses are given as
\beqa
M_i=-F_\phi\f{\al_i(\mu)}{4\pi}\(b_i-d\Delta b_i\)~,
\eeqa
with
\beqa
~(b_1~,b_2~,~b_3)&=&(\f{33}{5},~1,-3)~,~~\nn\\
\Delta(b_1~,b_2~,~b_3)&=&(~7,~7,~7).
\eeqa
\item  The trilinear soft terms will be determined by the superpotential after replacing canonical normalized superfields. They are given by\footnote{Although $Z_3$ invariant NMSSM is adopt in this scenario, we list the expressions of the most general GNMSSM spectrum. The $Z_3$ invariant results can be obtained by setting $\xi_S$ etc to vanish.}
\beqa
A_0^{ijk}\equiv \f{A_{ijk}}{y_{ijk}}&=&\sum\limits_{i}\(-\f{F_\phi}{2}\f{\pa}{\pa\ln\mu}+{d F_\phi}\f{\pa}{\pa\ln X}\) \ln \[Z_i(\mu,X,T)\]~,\nn\\
&=&\sum\limits_{i} \(-\f{F_\phi}{2} G_i^- +d F_\phi\f{\Delta G_i}{2}\)~.
\eeqa
Similarly, we can obtain the $m_{S^\pr}^2$ and $\xi_S$ terms.
The trilinear soft terms etc are given by
\beqa
A_t&=&\f{F_\phi}{16\pi^2}\[\tl{G}_{y_t}-2d \(y^u_{\bf 15}\)^2\]~,\nn\\
A_b&=&\f{F_\phi}{16\pi^2}\left\{\tl{G}_{y_b}-d\left[ 5\(y^L_{\bf 15;3}\)^2+2\(y^d_{\bf 15}\)^2\right]\right\}~,\nn\\
A_\tau&=&\f{F_\phi}{16\pi^2}\left\{\tl{G}_{y_\tau}-d\left[ 5\(y^L_{\bf 15;3}\)^2+2\(y^d_{\bf 15}\)^2\right]\right\}~,\nn\\
A_\la &=&\f{F_\phi}{16\pi^2}\left\{\tl{G}_{\la}-d\[2\(y^u_{\bf 15}\)^2+2\(y^d_{\bf 15}\)^2\]\right\}~,\nn\\
A_\ka &=&\f{F_\phi}{16\pi^2}\[\tl{G}_{\ka}\]~,\nn\\
m_{S^\pr}^2&=&{\mu^\pr}\f{F_\phi}{16\pi^2} \f{2}{3}\tl{G}_{\ka}~,\nn\\
\xi_S&=&{\xi_F}\f{F_\phi}{16\pi^2} \f{1}{3}\tl{G}_{\ka}~,\nn\\
m_3^2&=&{\mu}\f{F_\phi}{16\pi^2} \left\{\tl{G}_{H_u,H_d}-d\[2\(y^u_{\bf 15}\)^2+2\(y^d_{\bf 15}\)^2\]\right\}~,
\label{AMSB:trilinear}
\eeqa
with
\beqa
\tl{G}_{\la}&=&4\la^2+2\ka^2+3y_t^2+3y_b^2-(3g_2^2+\f{3}{5}g_1^2)~,\nn\\
\tl{G}_{\ka}&=&6\la^2+6\ka^2~,\nn\\
\tl{G}_{y_t}&=&\la^2+6y_t^2+y_b^2-(\f{16}{3}g_3^2+3g_2^2+\f{13}{15}g_1^2)~,\nn\\
\tl{G}_{y_b}&=&\la^2+y_t^2+6y_b^2-(\f{16}{3}g_3^2+3g_2^2+\f{7}{15}g_1^2)~,\nn\\
\tl{G}_{y_\tau}&=&\la^2+3y_b^2-(3g_2^2+\f{9}{5}g_1^2)~,\nn\\
\tl{G}_{H_u,H_d}&=&2\la^2+3y_t^2+3y_b^2-(3g_2^2+\f{3}{5}g_1^2)~.
\eeqa
\item The soft scalar masses are given by
\beqa
m^2_{soft}&=&-\left|-\f{F_\phi}{2}\f{\pa}{\pa\ln\mu}+d F_\phi\f{\pa}{\pa\ln X}\right|^2 \ln \[Z_i(\mu,X,T)\]~,\\
&=&-\(\f{F_\phi^2}{4}\f{\pa^2}{\pa (\ln\mu)^2}+\f{d^2F^2_\phi}{4}\f{\pa}{\pa(\ln |X|)^2}
-\f{d F^2_\phi}{2}\f{\pa^2}{\pa\ln|X|\pa\ln\mu}\) \ln \[Z_i(\mu,X,T)\],\nn
\eeqa
 Details of the expression involving the derivative of $\ln X$ can be found in our previous works~\cite{Fei:1508.01299, Fei:1602.01699,Fei:1703.10894}.

Expressions for scalars can be parameterized as the sum of each contributions
\beqa
m_{soft}^2=\delta_{A}+\delta_{I}+\delta_G~,
\label{AMSB:scalar}
\eeqa
with $\delta_{A}$ the anomaly mediation contributions, $\delta_G$ the general gauge(Yukawa) mediation contributions and $\delta_I$ the interference contributions, respectively. Because of each term is rather lengthy, we collect their expressions in appendix \ref{appendix:B}.
\eit
If the $y^d_\Delta \Delta_T H_d H_d$ term is present in the superpotential of GNMSSM, the soft SUSY breaking trilinear term can give subleading contribution to Majorana neutrino mass via type-II seesaw mechanism. The trilinear scalar coupling $\tl{\Delta}_T-H_d-H_d$ can be obtained before we integrate out the messengers involving $\Delta_T$
\beqa
A_{H_d H_d \Delta_T}=\f{y_{H_d H_d \Delta_T}}{2}\f{F_\phi}{16\pi^2}G_{H_d H_d \Delta_T}~,
\eeqa
with
\beqa
G_{H_d H_d \Delta_T}=y_X^2+ \sum\limits_{c}\(y^L_{\bf 15;c}\)^2+5 \(y^d_{\bf 15}\)^2+6y_b^2-\({7}g_2^2+\f{9}{5}g_1^2\),
\eeqa
the corresponding Yukawa beta function between $M_{mess}$ and $M_{GUT}$. In our scenario with $Z_3$ invariant NMSSM, such trilinear term vanishes because of vanishing $y^d_\Delta \Delta_T H_d H_d$ term.

\subsection{Numerical constraints on type-II neutrino seesaw mechanism extension of NMSSM from dAMSB}

  The free parameters for our economical type-II neutrino seesaw mechanism extension of NMSSM from deflected AMSB are given as
 \beqa
 F_\phi,~~~M_{mess},~~~d, ~~y^L_{\bf 15;a},~~y^d_{\bf 15},~~~y^u_{\bf 15},~~~\la,~~~\ka.
 \eeqa
 The spurion $X$ is also normalized so that $y_X=1$.
We also adopt
\beqas
y^L_{\bf 15;a}=\la_0~,~y^d_{\bf 15}=y^u_{\bf 15}=\la_1~,
\eeqas
to reduce further the free parameters of this scenario.

In ordinary  AMSB realization of NMSSM, large $A_\la,A_\ka$ needs large $\la$ and $\ka$ so as to induce large positive $m^2_S$, suppressing the singlet VEV~\cite{NMSSM:AMSB}. In our scenario, as can be seen in eqn(\ref{AMSB:trilinear}) and eqn(\ref{AMSB:scalar}), new interactions involving $H_u,H_d$ and triplets will lead to additional contributions to $A_\la$, possibly ameliorating the previous difficulties.

    We still use NMSSMTools 5.5.0~\cite{NMSSMTools} to scan the whole parameter space. Randomly scan in combine with MCMC method are used. Similar to the choice in GMSB, the range of the free parameters are chosen as
 \beqa
&&~~ 10^{11} ~{\rm GeV}< M_{mess}< 2.0\tm 10^{14}~{\rm GeV}~,~~ 10 ~{\rm TeV} <F_\phi ~< 500 ~{\rm TeV}~,~~-5<d<5,\nn\\
&&~~~~~~~~~~~0<\la_0,\la_1<\sqrt{4\pi}~,~~~~~~0.1<\la,\ka <0.7~~ {\rm and}~~~ \la^2+\ka^2\lesssim 0.7~.
 \eeqa
 Bounds for $\la_1$ and $M_{mess}$ from LFV are also similar to that of GMSB.
 In our scan, in addition to the constraints of the neutrino masses eqn(\ref{neutrino:part}), constraints from (I) to (IV) in GMSB scenario are also imposed here. Besides, we also impose the following constraints
  \bit
  \item (V) The purpose of deflection in AMSB is to solve the notorious tachyonic slepton problem. So non-tachyonic sleptons should be obtained after RGE running to the SUSY scale.
  \item (VI) The relic density of cosmic DM should satisfy the Planck data $\Omega_{DM} = 0.1199\pm 0.0027$ ~\cite{Planck} in combination with the WMAP data ~\cite{WMAP}(with a $10\%$ theoretical uncertainty). We impose only the upper bound of $\Omega_{DM}$ in our numerical studies because other DM species can also possibly contribute to the relic abundance of DM.
\eit
   In NMSSM, the 125 GeV Higgs boson in general can be either the lightest or the next-to-lightest CP-even scalar. Depending on the nature of the 125 GeV Higgs, we have the following discussions related to our numerical results
    \bit
    \item {\bf A})  125 GeV Higgs is the lightest CP-even scalar.

\begin{figure}[htb]
\begin{center}
\includegraphics[width=2.9in]{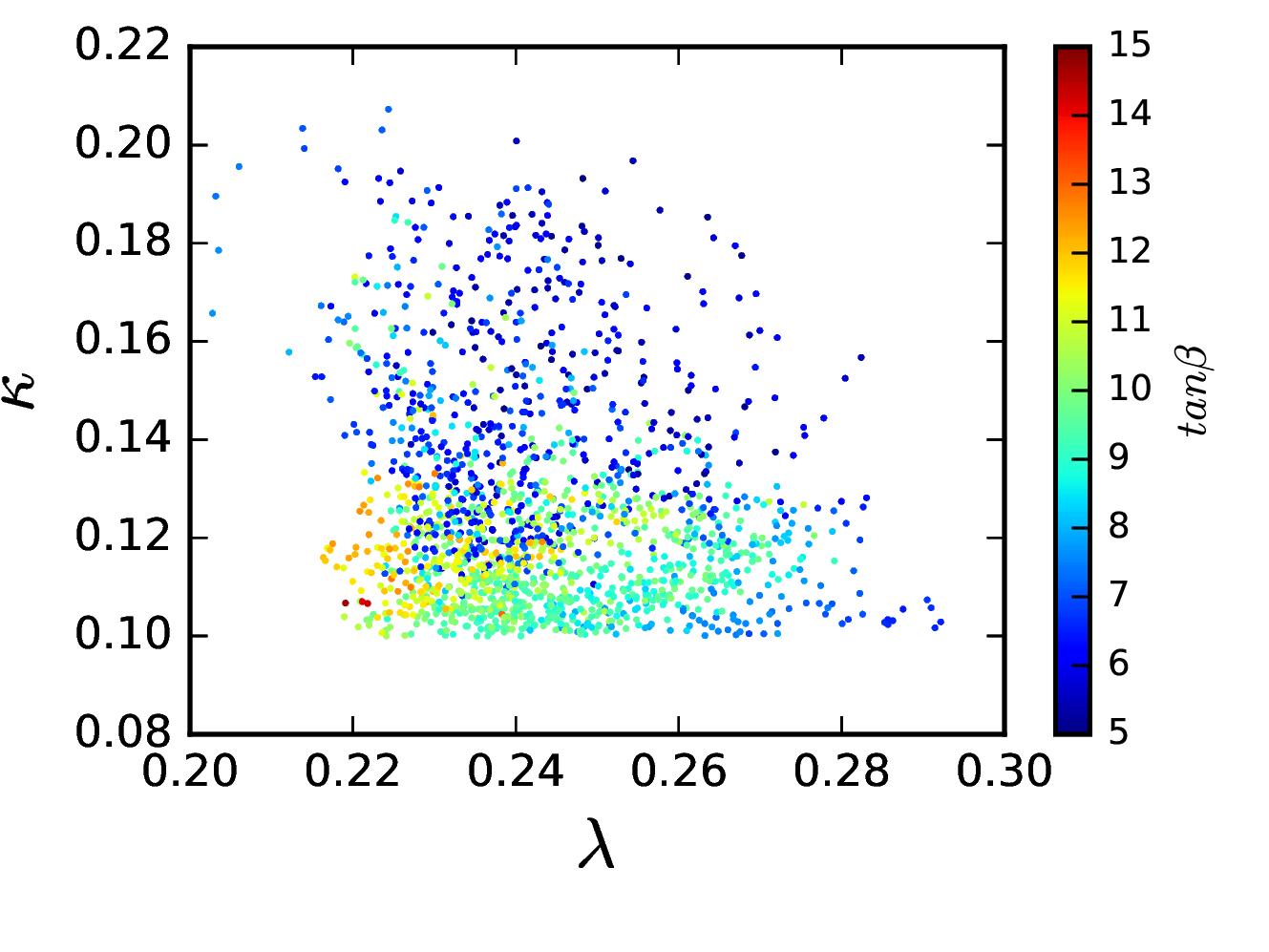}
\includegraphics[width=2.9in]{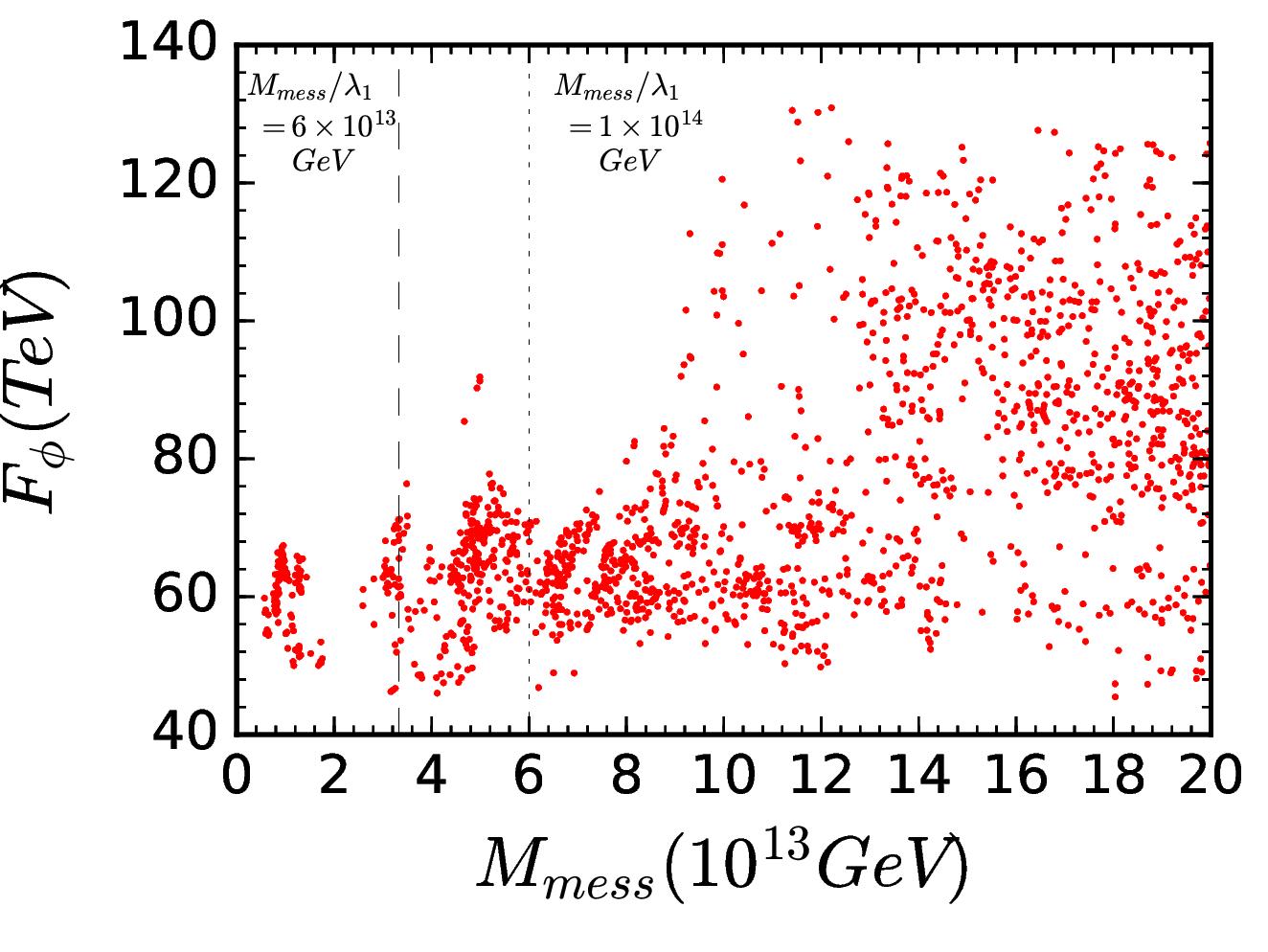}\\
\includegraphics[width=2.9in]{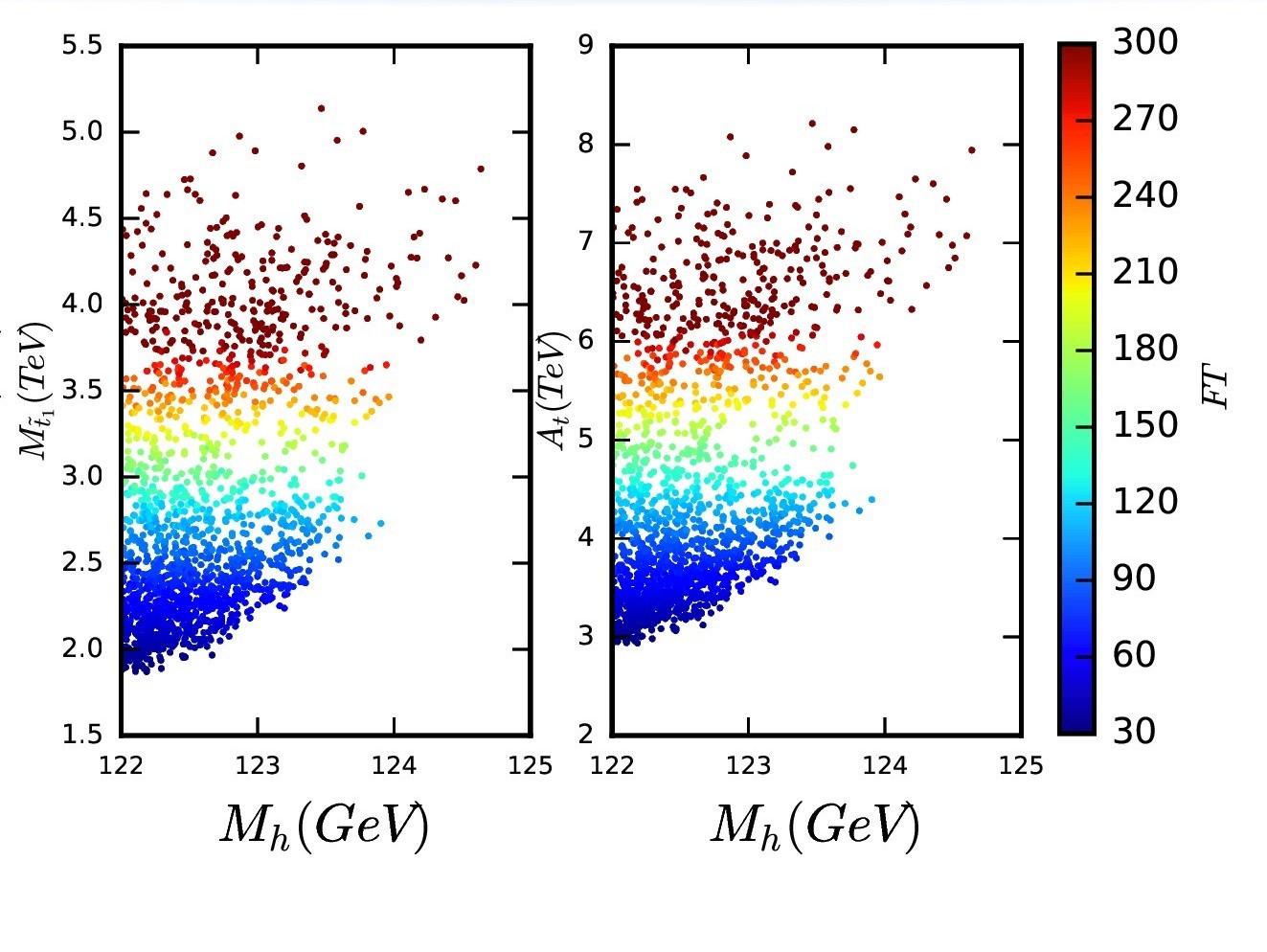}
\includegraphics[width=2.9in]{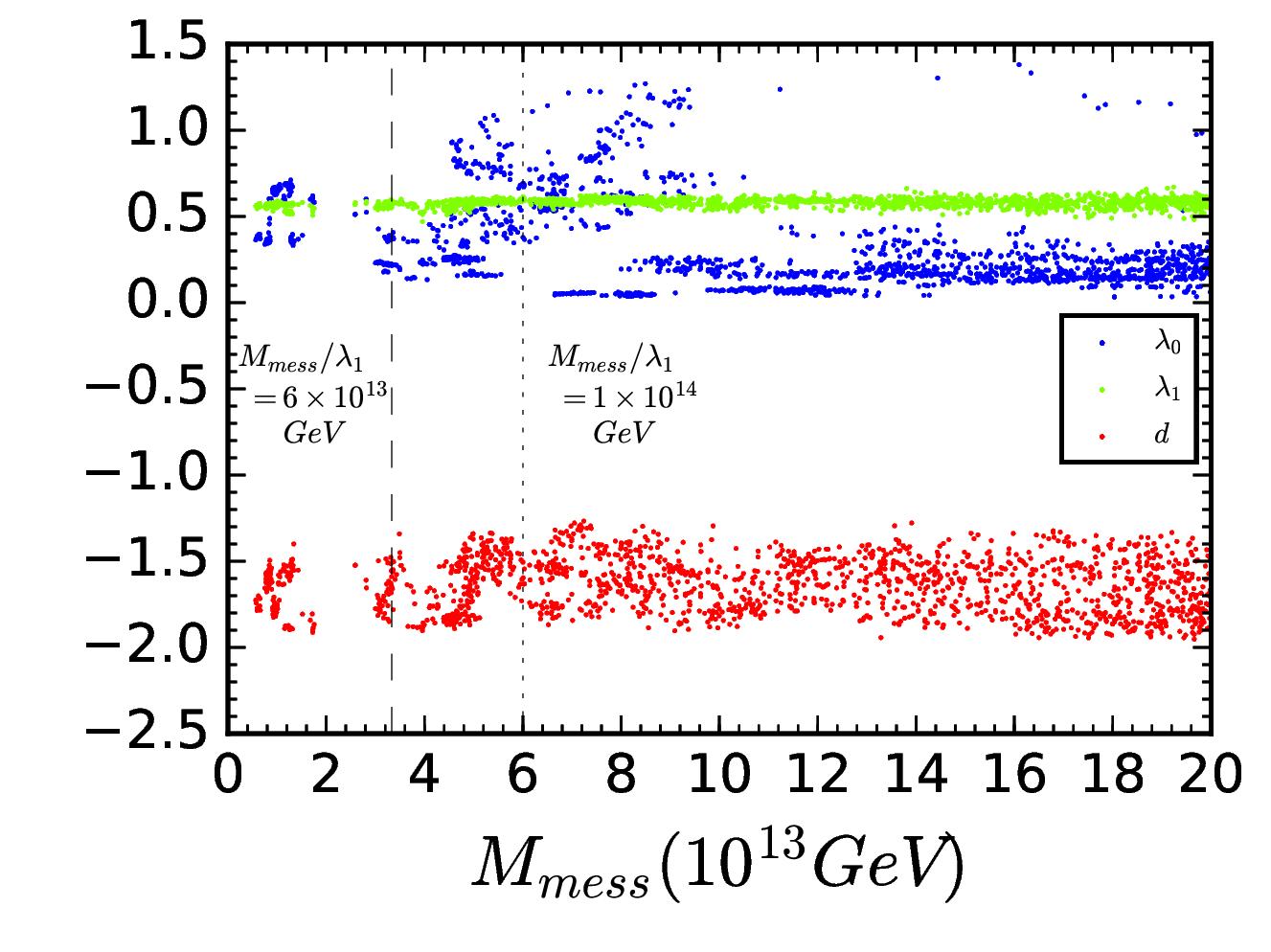}\\
\includegraphics[width=2.9in]{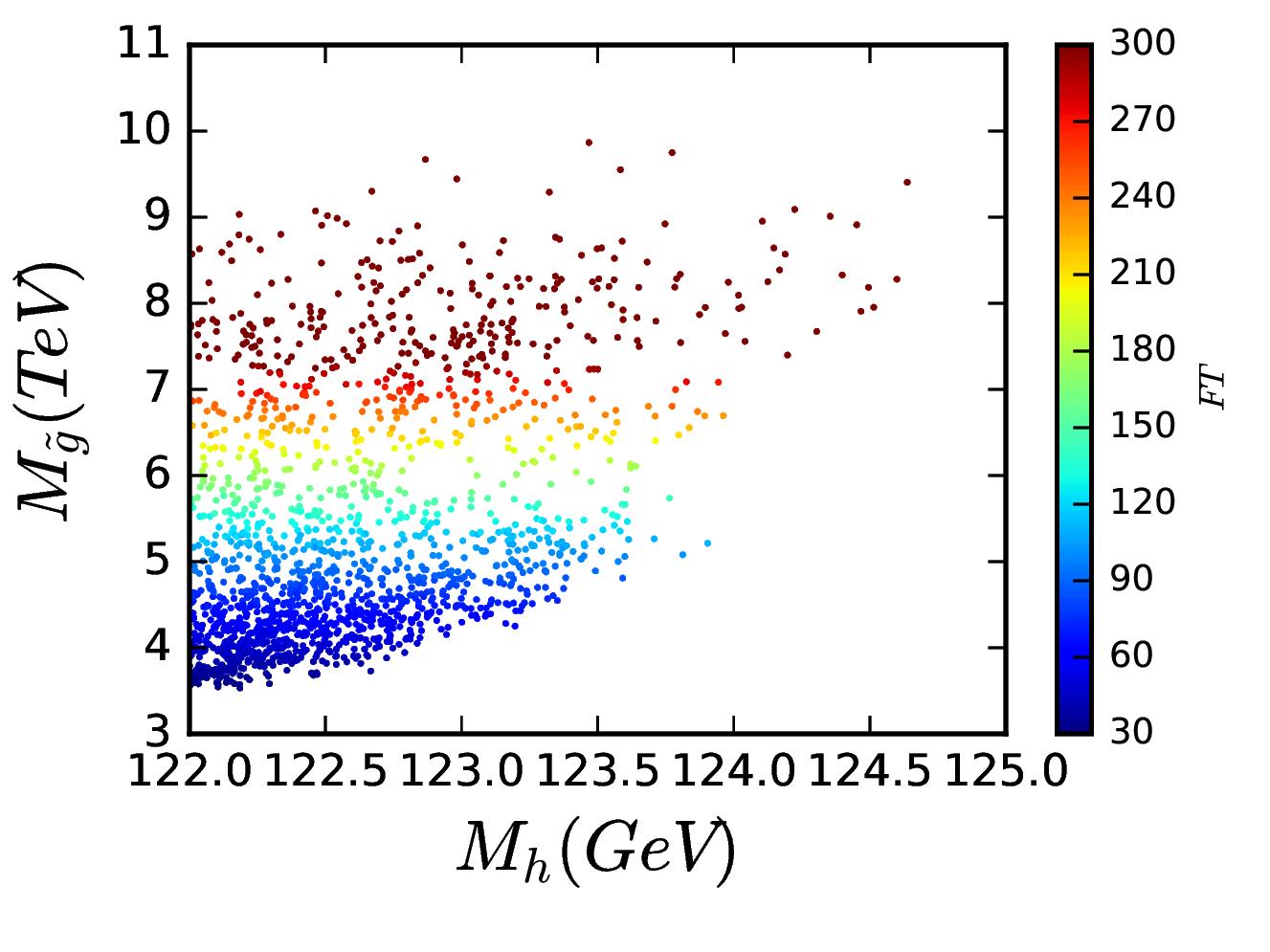}
\includegraphics[width=2.9in]{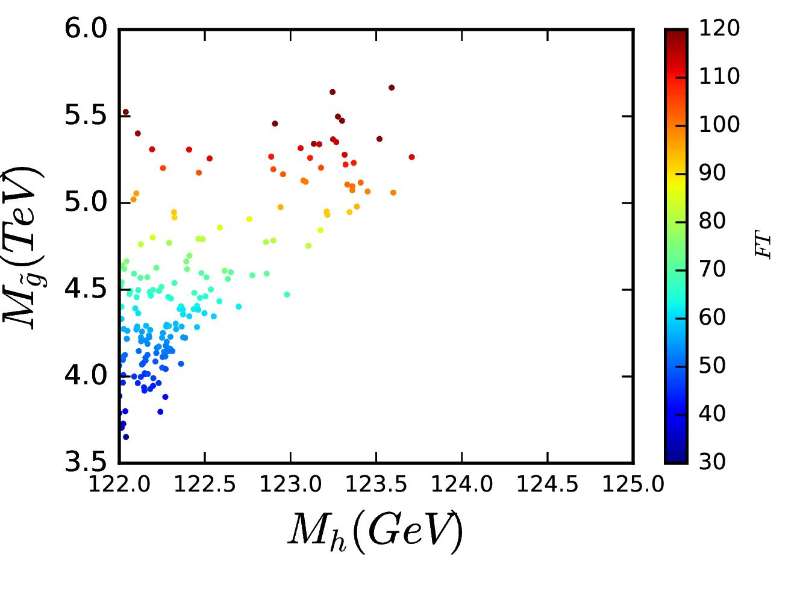}\\
\end{center}
\vspace{-.5cm}
\caption{ Survived points that can satisfy the EWSB conditions and the constraints from (I) to (V) in case the 125 GeV Higgs is the lightest CP-even scalar in AMSB-type scenario. The lower right panel shows the survived points with additional LFV bound $M_{mess}/\la_1\lesssim (0.6\tm 10^{14}{\rm GeV})$.}
\label{fig2}
\end{figure}


         Our numerical scan indicates that EWSB conditions alone can already ruled out a large portion of the total parameter space. Combing with the constraints from (I) to (VI), we can obtain the survived points that lead to realistic SUSY spectrum at low energy, which are shown in fig.\ref{fig2}. In the left panel of fig.\ref{fig2}, the allowed $\ka$ versus $\la$ regions are given. We can see that $\ka$ should lie between $0.1$ to $0.24$ while $\la$ should lie between $0.19$ to $0.29$. The $\tan\beta$, which is obtained iteratively from the minimization condition of the Higgs potential for EWSB, are constrained to lie between $4$ and $14$ for $40~{\rm TeV}\leq F_\phi\leq 140~{\rm TeV}$.

         It is interesting to note that the messenger scale, which is just the heavy triplet scalar scale in type-II seesaw mechanism, are constrained to lie in a small band, from $0.6\tm 10^{13}$ GeV to $2.0\tm 10^{14}$ GeV. Lower values of $M_{mess}$ are ruled out by Higgs masses and LHC data. 
         The allowed ranges of $\la_1$, which is just the coupling $y_\Delta^u$, can be seen to lie in a very narrow band centered at $\la_1\approx 0.55$. Neutrino masses bounds alone allow light $M_{mess}$ with tiny $y^u_\Delta$. However, successful EWSB in NMSSM as well as non-tachyonic slepton requirements etc forbid too small $y^u_\Delta$, as relatively large couplings are needed to give non-negligible Yukawa mediation contributions to the soft SUSY breaking parameters. We also show the possible exclusion lines from $l_i\ra l_j \ga$ LFV processes, which give an upper bounds for $M_{mess}/\la_1$. For example, the conservative requirement $M_{mess}/\la_1\lesssim (0.6\tm 10^{14}{\rm GeV})$ will set an upper bound for messenger scale to be $3.3\tm 10^{13}{\rm GeV}$. We left the detailed discussions on LFV bounds in AMSB scenarios in our future works.

     The plot of Higgs mass $m_h$ versus $A_t$ or lighter stop mass $\tl{t}_1$ are shown in the middle left panel of fig.\ref{fig2}. It is clear from the panel that our scenario can successfully account for the 125 GeV Higgs boson. From the allowed values of $\la$ and $\tan\beta$, it can be seen that the NMSSM specific $\la^2 v^2 \sin^22\beta$ contributions to the Higgs mass $m_h^2$ is small, which is estimate to be $49 ~{\rm GeV}^2$ for $\tan\beta=10$. We note that this small contribution to Higgs mass is still much larger than the case of GMSB. So, large $A_t$ or heavy stop is necessary to give the 125 GeV Higgs. Fortunately,  $A_t$ receives additional contributions in our scenario, which will increase $A_t$ for negative deflection parameters. Besides, unlike our GMSB case, the ratio $A_t/m_{\tl{T}}$ lies much nearer to the maximal mixing value $A_t/m_{\tl{T}}\simeq \pm\sqrt{6}$, making the second term of the second line of eqn.(\ref{higgs:NMSSM}) to give important contributions to Higgs mass.  So the radiative correction from $\ln(m_{\tl{T}}/m_t)$ term needs not be too large, making light $m_{\tl{T}}$ possible. As $m_{\tl{T}}$, which is typically determined by $F_\phi$, characterizes the mass scale for colored sparticles, the SUSY breaking spectrum of other colored sparticles can be relatively light.

The middle right panel of fig.\ref{fig2} shows that the preferred deflection parameters lie between $-1.3$ and $-2.0$, which indeed increase the value of $A_t$. It can also be seen from this panel that the allowed regions require non-vanishing $\la_0$,$\la_1$ couplings, which means that Yukawa deflection in AMSB by the triplet messengers etc is indispensable to obtain realistic low energy SUSY spectrum.  We can see that stop as light as almost 2 TeV can interpret the observed 125 GeV Higgs. It can also be seen that larger $A_t$ or $\tl{t}_1$ can predict larger Higgs mass as expected. The values of $F_\phi$ can determine the whole scales of the soft spectrum. We can see from the lower left panel of fig.\ref{fig2} that the gluino are constrained to lie upon 3.5 TeV. Such a heavy gluino can evade the current LHC bounds. The upper bounds for gluino is 10 TeV, which is not accessible in the near future experiment. We also show the allowed Higgs mass with additional LFV constraints $M_{mess}/\la_1\lesssim (0.6\tm 10^{14}{\rm GeV})$ in the lower right panel of fig.\ref{fig2}. We can see that the predicted Higgs mass can not exceed 124 GeV because lower messenger scale will lead to smaller RGE contributions to soft scalar masses. The gluino mass will also be bounded to be less than 5.8 TeV with this constraint.

 The Barbieri-Giudice FT measures of our scenario are shown in the middle left panel of fig.\ref{fig2}. In the allowed region, the BGFT satisfies $\Delta\lesssim 300$. The FT can be as low as 30 in the low gluino mass regions. We know that the gluino mass is determined by $F_\phi$, which set all the soft SUSY mass scale. So, lighter $m_{\tl{g}}$ in general indicates lighter stop. We can see that lighter $A_t$ or $\tl{t}_1$ will lead to smaller BGFT for fixed Higgs mass. On the other hand, increasing $\tl{t}_1$ while at the same time increasing $A_t$ can possibly make the involved FT unchanged. This conclusion agrees with conclusions from the electroweak FT measure $\Delta_{EW}$~\cite{rnaturalsusy}. As the BGFT in general will overestimate the FT involved, the region with intermediate BGFT can still be natural. To illustrate the spectrum of this scenario, we show a benchmark point in Table \ref{Delta1}.
\begin{table}[htbp]
\caption{Benchmark point for type-II neutrino seesaw mechanism extension of NMSSM from AMSB in the case that the 125 GeV Higgs is the lightest CP-even scalar. All mass parameters are in the unit of GeV. $\Delta a_\mu$ denotes additional SUSY contributions to muon anomalous magnetic momentum.}
 \begin{tabular}{|c|c|c|c|c|c|}
 \hline
 \hline
 ${F_{\phi}}$&$51755.860$&$M_{mess}$&$0.152\tm10^{14}$&$\la_0$&$0.391$\\
 \hline
$\la_1$&$0.581$&$d$&$-1.807$&$\la$&$0.264$\\
 \hline
 $\ka$&$0.110$&&&&\\
 \hline\hline
 $\tan{\beta}$&$7.690$&$A_\la$&$1228.001$&$A_\ka$&$-78.285$\\
 \hline
 $A_t$&$3472.252$&$A_b$&$-634.383$&$A_\tau$&$1468.157$\\
 \hline
 $m_{h_1}$&$122.013$&$m_{h_2}$&$136.703$&$m_{h_3}$&$1392.932$\\
 \hline
 $m_{a_1}$&$132.426$&$m_{a_2}$&$31392.498$&$m_h^{\pm}$&$1393.401$\\
 \hline
 $m_{\tl{d}_{L}}$&$2668.521$&$m_{\tl{d}_{R}}$&$2364.470$&$m_{\tl{u}_{L}}$&$2667.408$\\
 \hline
 $m_{\tl{u}_{R}}$&$2388.512$&$m_{\tl{s}_{L}}$&$2668.521$&$m_{\tl{s}_{R}}$&$2364.470$\\
 \hline
$m_{\tl{c}_{L}}$&$2667.408$&$m_{\tl{c}_{R}}$&$2388.512$&$m_{\tl{b}_{1}}$&$2287.543$\\
 \hline
$m_{\tl{b}_{2}}$&$2583.751$&$m_{\tl{t}_{1}}$&$2289.537$&$m_{\tl{t}_{2}}$&$2613.992$\\
 \hline
$m_{\tl{e}_{L}}$&$1389.267$&$m_{\tl{e}_{R}}$&$812.087$&$m_{\tl{\nu_e}}$&$1387.203$\\
 \hline
$m_{\tl{\mu}_{L}}$&$1389.267$&$m_{\tl{\mu}_{R}}$&$812.087$&$m_{\tl{\nu_{\mu}}}$&$1387.203$\\
 \hline
$m_{\tl{\tau}_{1}}$&$804.858$&$m_{\tl{\tau}_{2}}$&$1387.217$&$m_{\tl{\nu_{\tau}}}$&$1385.149$\\
 \hline
$m_{\tl{\chi}_{1}^{0}}$&$147.658$&$m_{\tl{\chi}_{2}^{0}}$&$-194.412$&$m_{\tl{\chi}_{3}^{0}}$&$211.456$\\
 \hline
 $m_{\tl{\chi}_{4}^{0}}$&$-1324.742$&$m_{\tl{\chi}_{5}^{0}}$&$-1855.360$&$m_{\tl{\chi}_{1}^{\pm}}$&$193.185$\\
 \hline
 $m_{\tl{\chi}_{2}^{\pm}}$&$-1855.356$&$\mu_{eff}$&$185.51$&$m_{\tl{g}}$&$4625.582$\\
 \hline
$\Omega_\chi h^2$&$0.038$&$\sigma_P^{SI}$&$0.205\tm10^{-10}pb$&$\Delta a_\mu$&$-5.622\tm 10^{-11}$\\
 \hline
 \hline
 \end{tabular}
\label{Delta1}
\end{table}

\begin{figure}[htb]
\begin{center}
\includegraphics[width=2.9in]{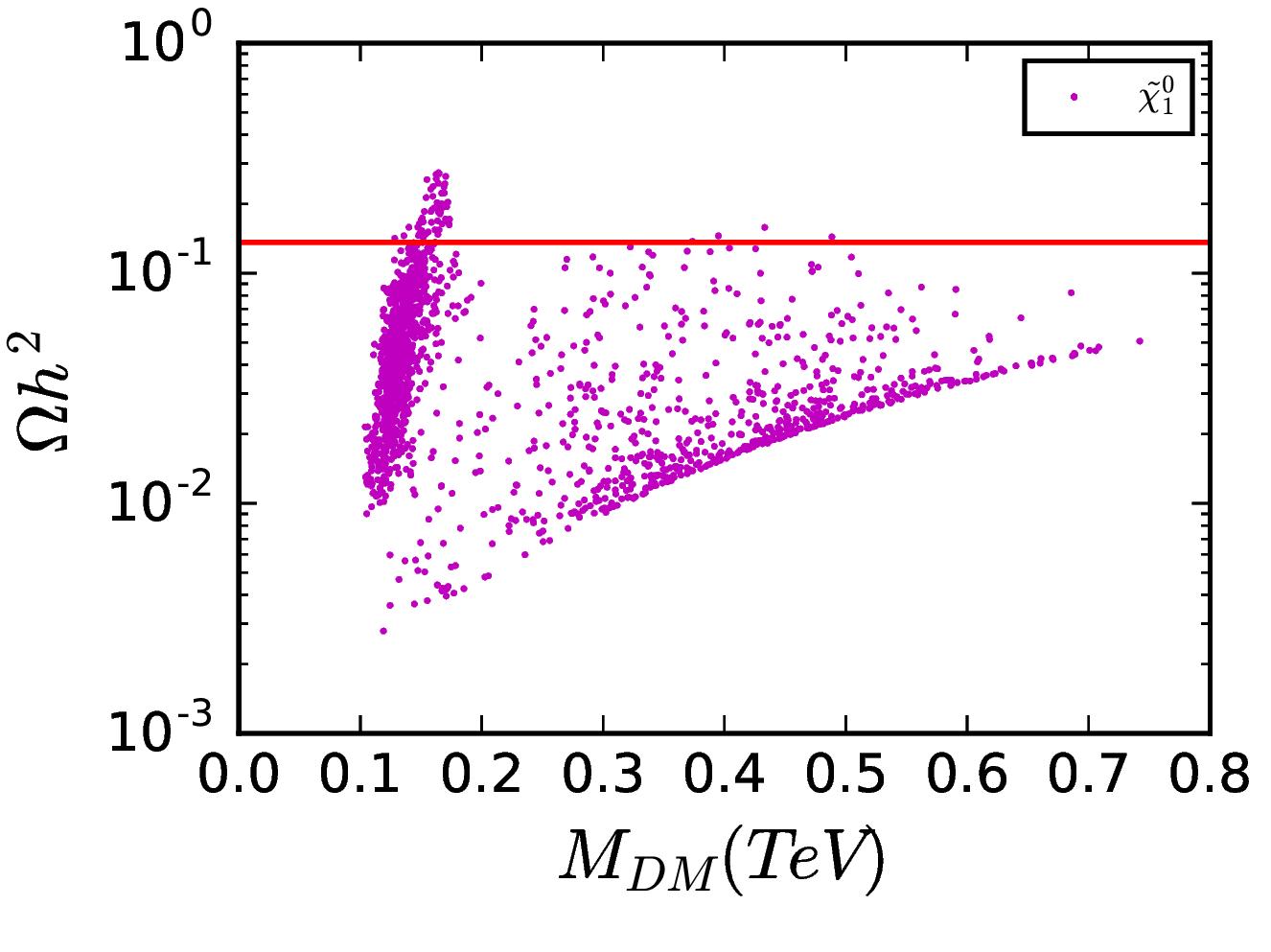}
\includegraphics[width=2.9in]{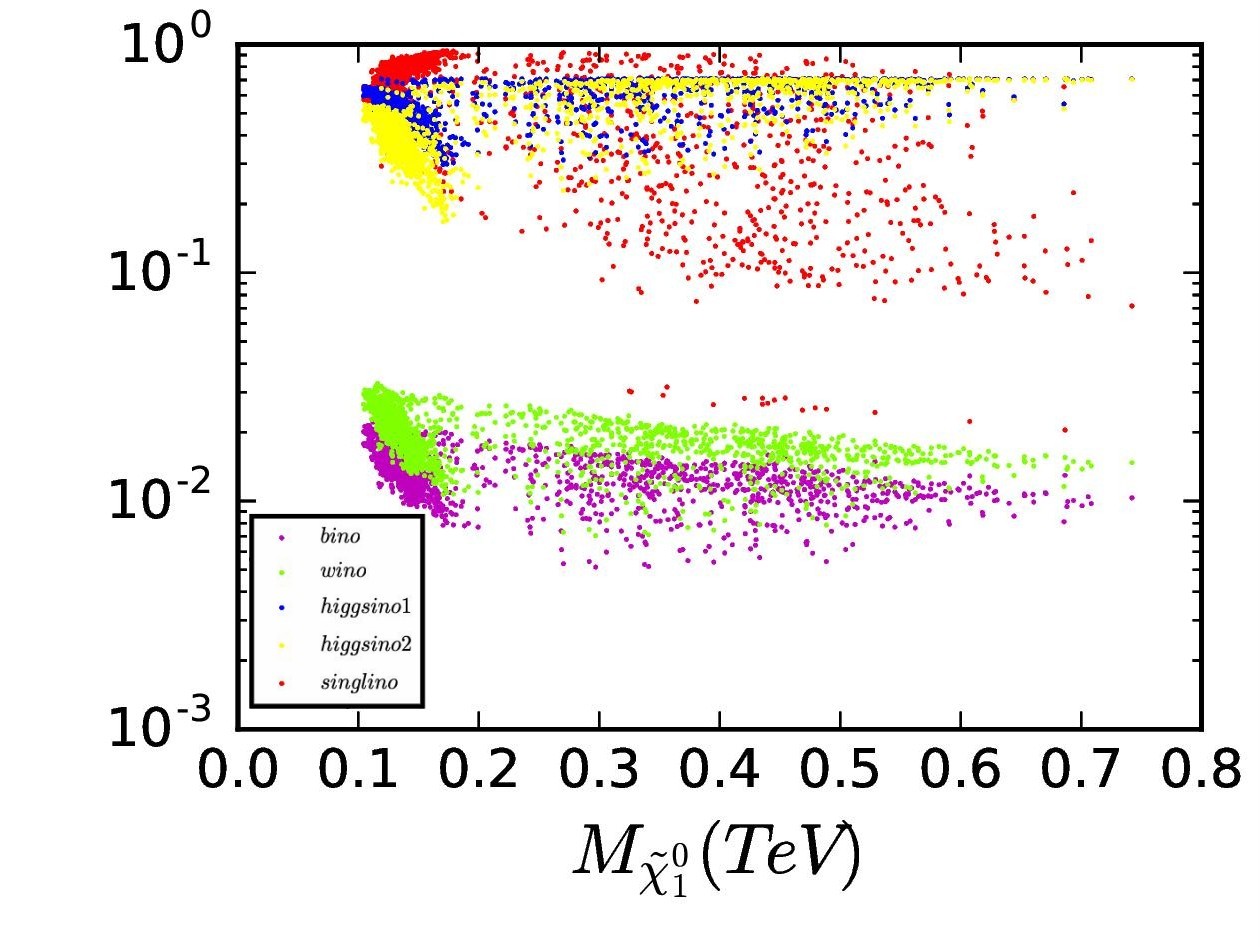}\\
\includegraphics[width=2.9in]{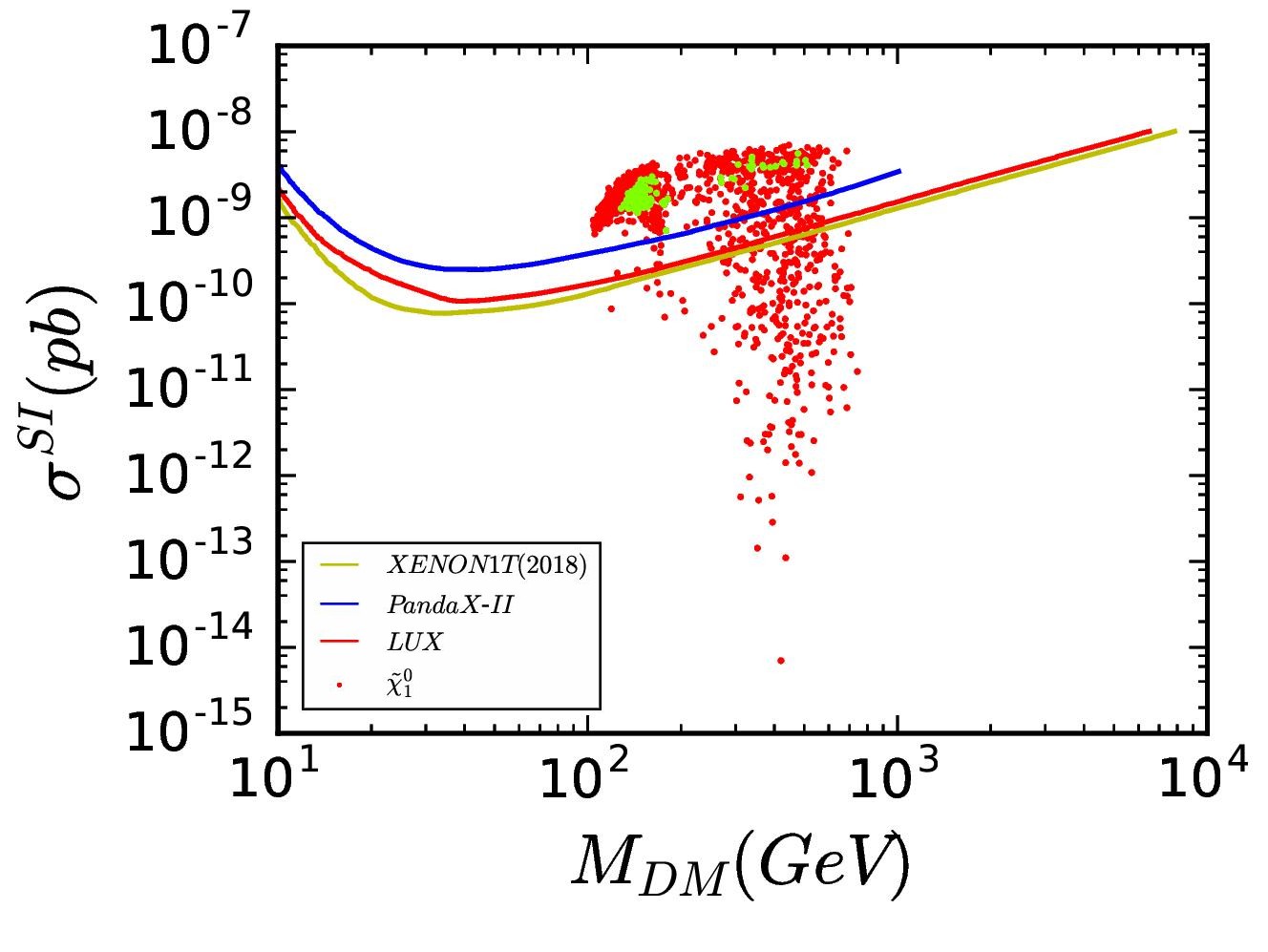}
\includegraphics[width=2.9in]{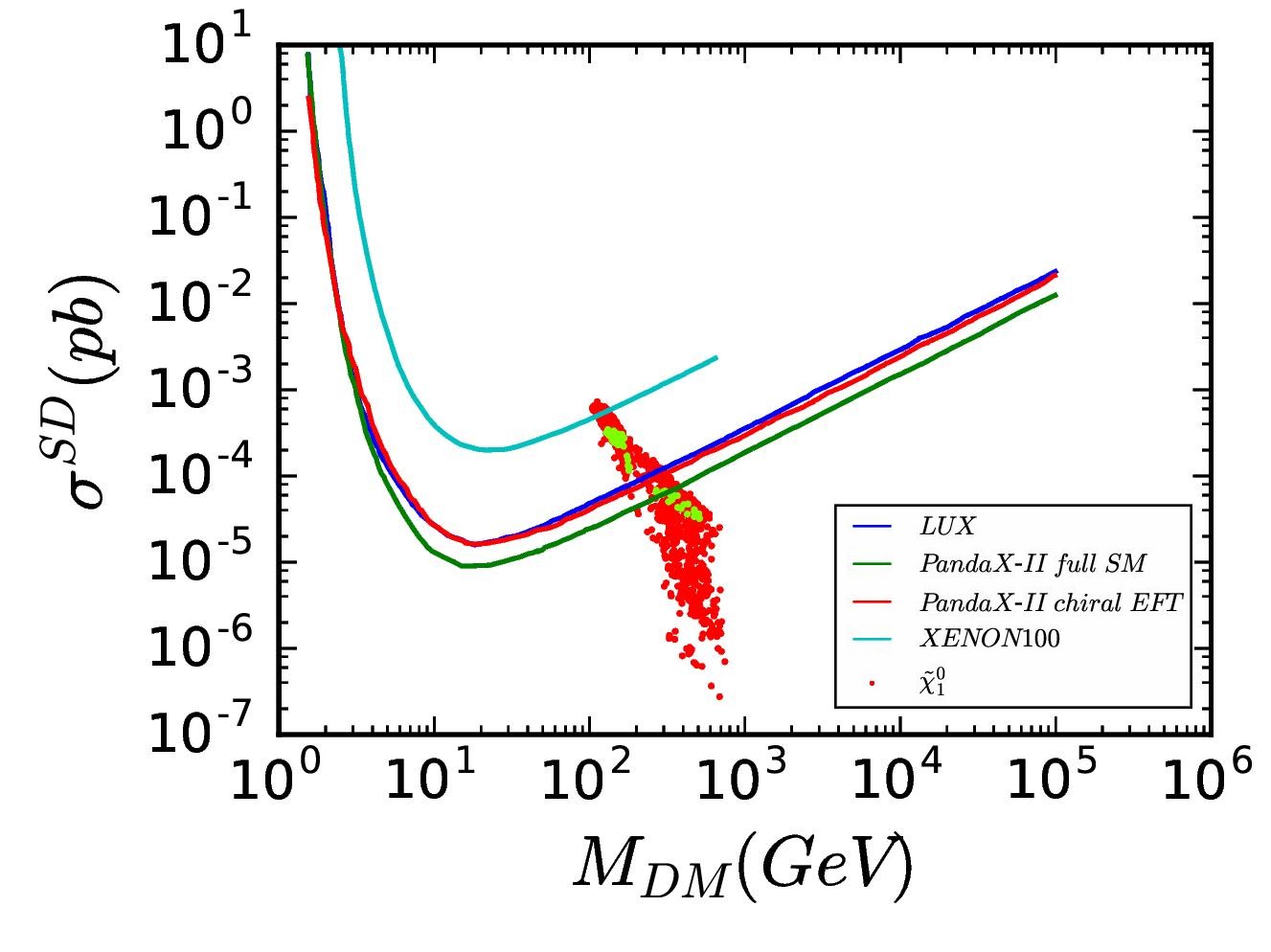}\\
\end{center}
\vspace{-.5cm}
\caption{  Relic density of the neutralino DM versus the DM mass is given in the upper left panel. The ingredients of the neutralino DM is shown in the upper right panel. The  spin-independent cross section $\sigma_{SI}$ (left) and the spin-dependent cross section $\sigma_{SD}$ (right) versus DM mass for DM direct detection experiments are given in the lower panels, respectively. The green points denote the parameters that can provide full DM relic abundances.
}
\label{fig3}
\end{figure}

 In AMSB, the gravitino mass will be of order $F_\phi$, which is heavier than ordinary soft SUSY breaking parameters. Therefore, the lightest neutralino can act as the DM candidate. We can see from the upper left panel of fig.\ref{fig3} that in most of the previous allowed parameter space, the neutralino will lead to under-abundance of DM, although full abundance of DM is still possible for a small portion of the parameter space. This can be understood from the ingredients of the neutralino, which is shown in the upper right panel. We can see that DM is singlino-dominant in most of the parameter space.
The almost pure singlino-like DM is a distinctive feature of NMSSM. Its relic density can be compatible with WMAP bounds if it can annihilate via s-channel CP-even (or CP-odd) Higgs exchange when such Higgs  has sufficient large singlet component for not too small $\kappa$. Under abundance of DM will not cause a problem as other specie of DM, such as axion or axino, can possibly contribute to the remaining abundances of DM.

The direct detection experiments, such as LUX~\cite{LUX2016},Xenon~\cite{XENON1T2018},PandaX~\cite{PANDAX}, will set upper limits on the WIMP-nucleon scattering cross section. The spin-independent(SI) and spin-dependent(SD) scattering cross section of the neutralino DM is displayed in the left and right lower panels in fig.\ref{fig3}, respectively. As the SI neutralino-nucleon interaction arises from s-channel squark, t-channel Higgs (or Z) exchange at the tree level and neutralino-gluon interactions from the one-loop level involving quark loops, singlino-like DM can evade the direct detection constraints if the t-channel exchanged Higgs is not too light for heavy squarks. The SD neutralino-nucleon interaction is dominated by $Z^0$ exchange for heavy squarks with the corresponding cross section proportional to the difference of the Higgsino components $\si_{SD}\propto |N_{13}^2-N_{14}^2|$. If the two Higgsino components are large but similar, the SD cross section can become small, which however will lead to large  $\si_{SI}$ as $\si_{SI}\propto |N_{13}^2+N_{14}^2|$. We can see from the middle panels that although some portion of the allowed parameter space is ruled out by DM direct detection experiments, especially by $\si_{SI}$ in case the singlino-like DM provides full abundance of DM(the green points), a large portion of parameter space is still not reached by current experiments if there are other DM components other than the lightest neutralino.

  \item {\bf B}) The 125 GeV Higgs is the next-to-lightest CP-even scalar.

\begin{figure}
\begin{center}
\includegraphics[width=2.8in]{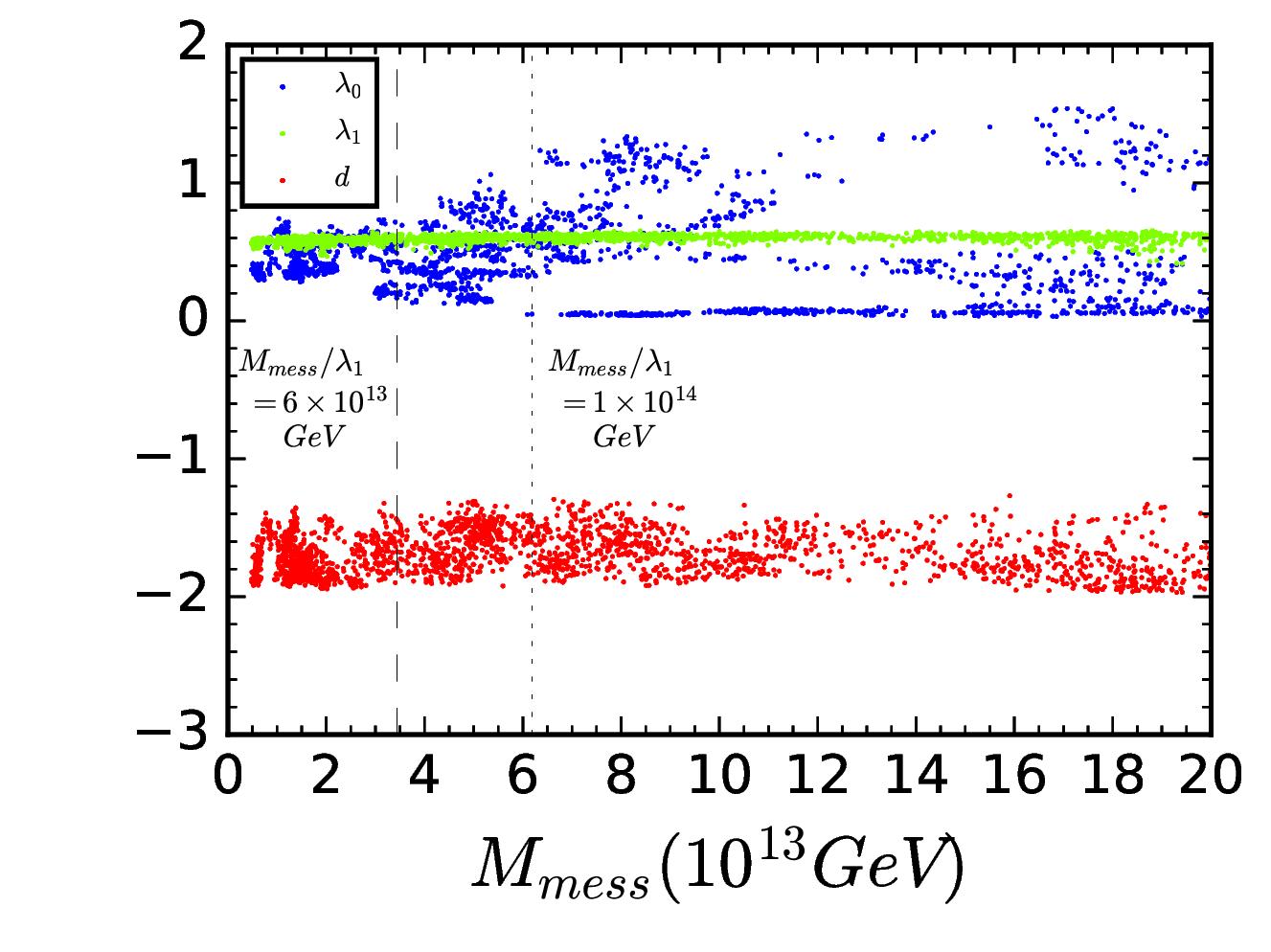}
\includegraphics[width=2.8in]{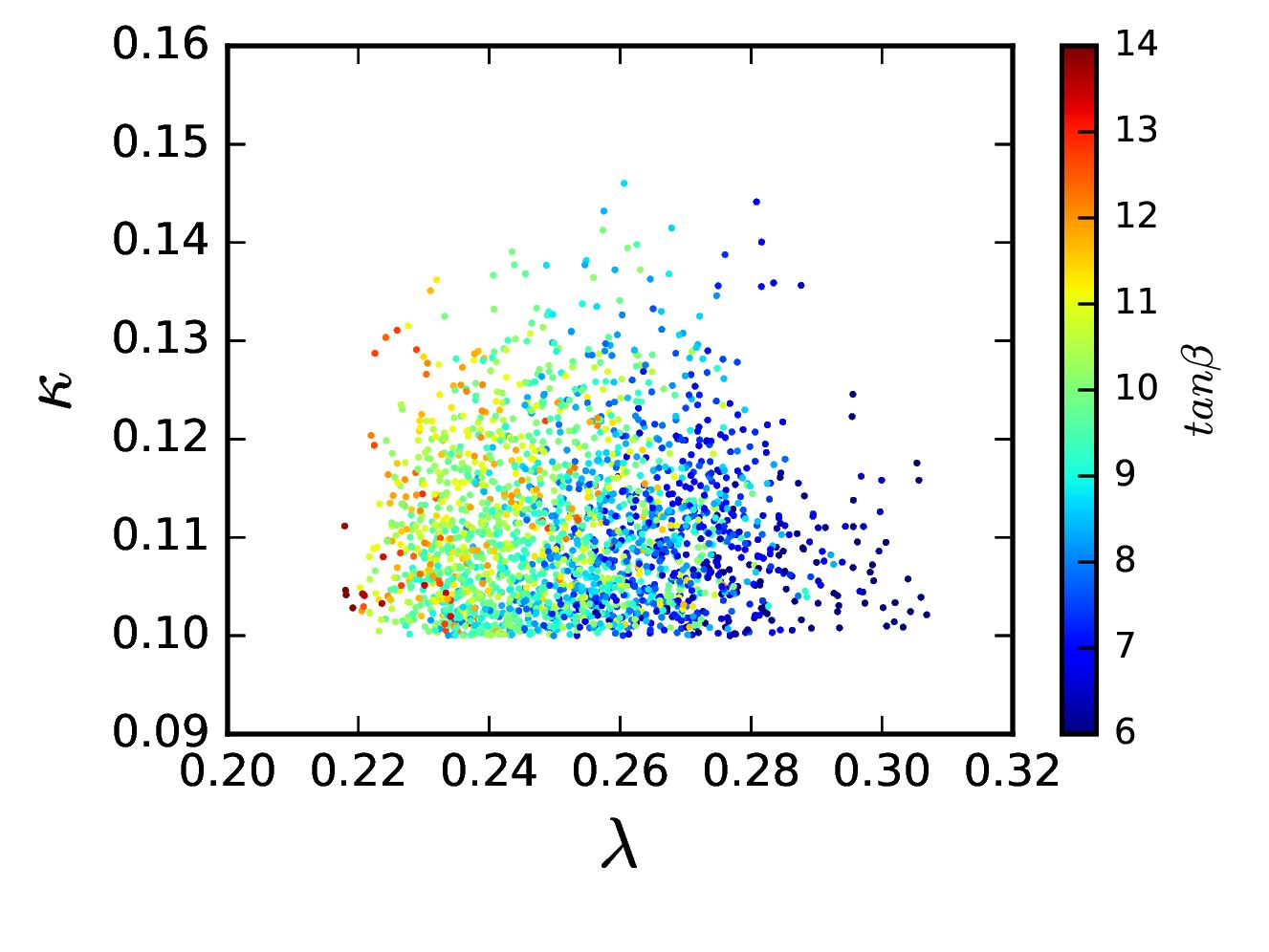}\\
\includegraphics[width=2.8in]{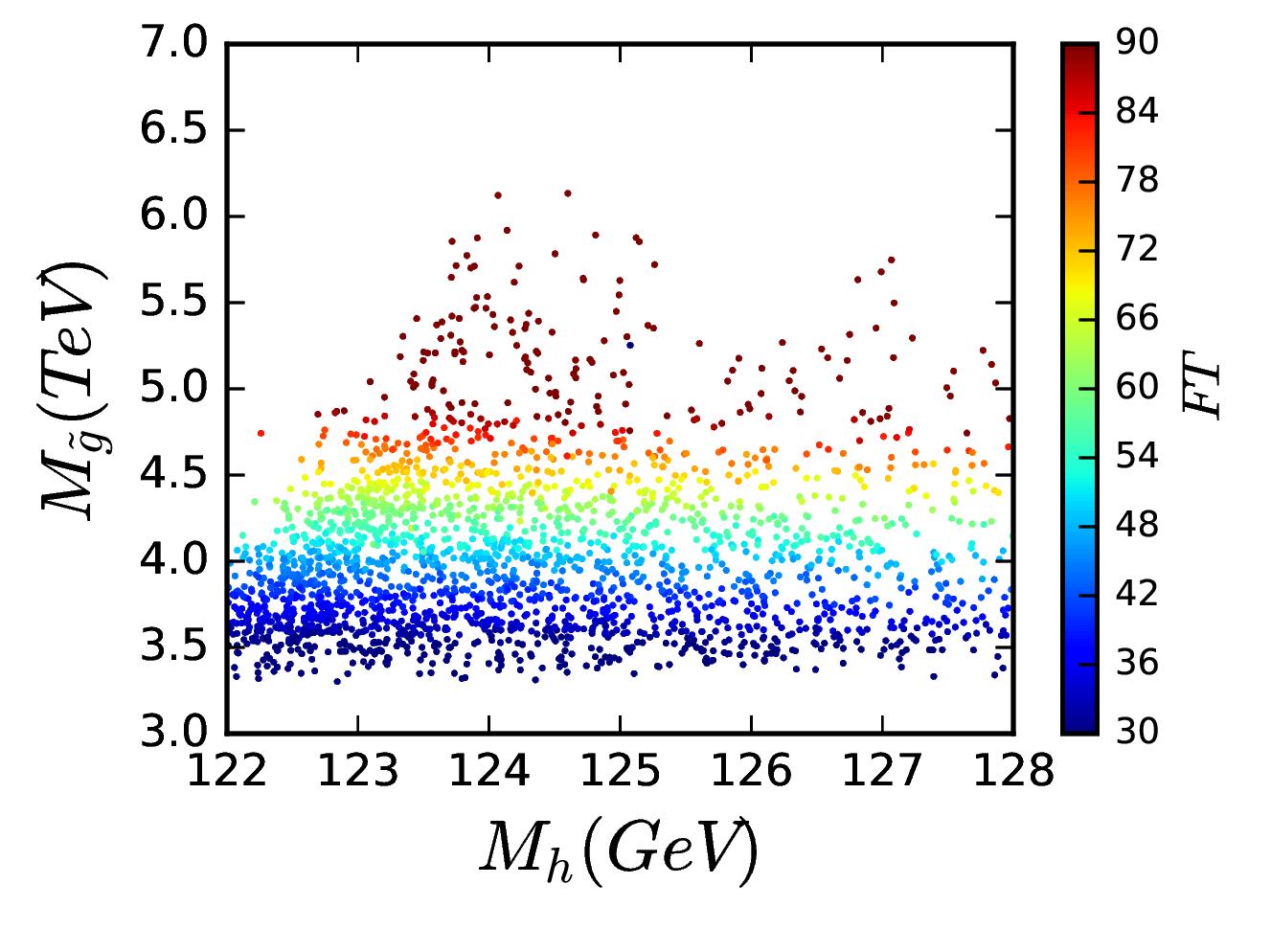}
\includegraphics[width=2.8in]{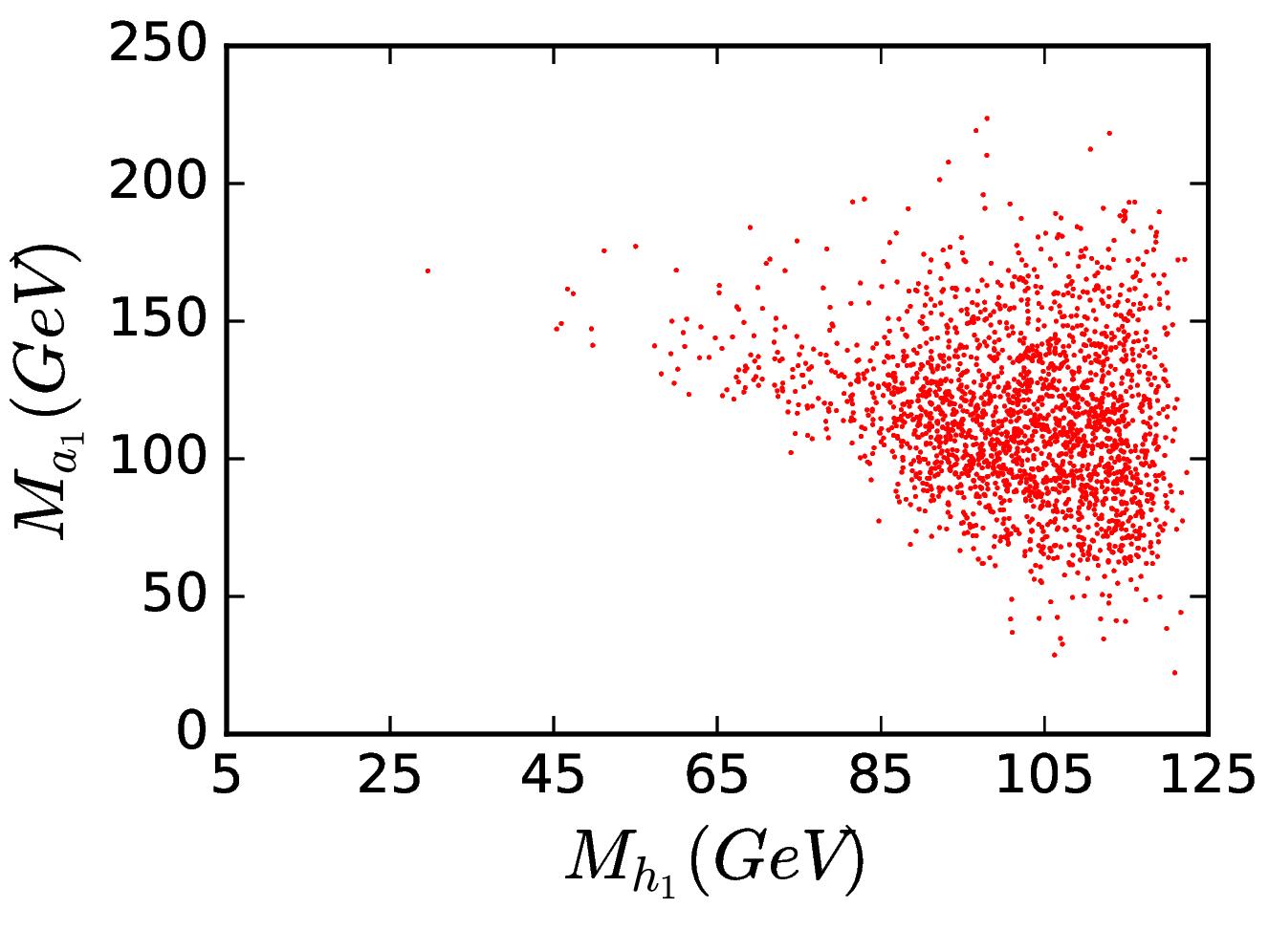}\\
\includegraphics[width=2.8in]{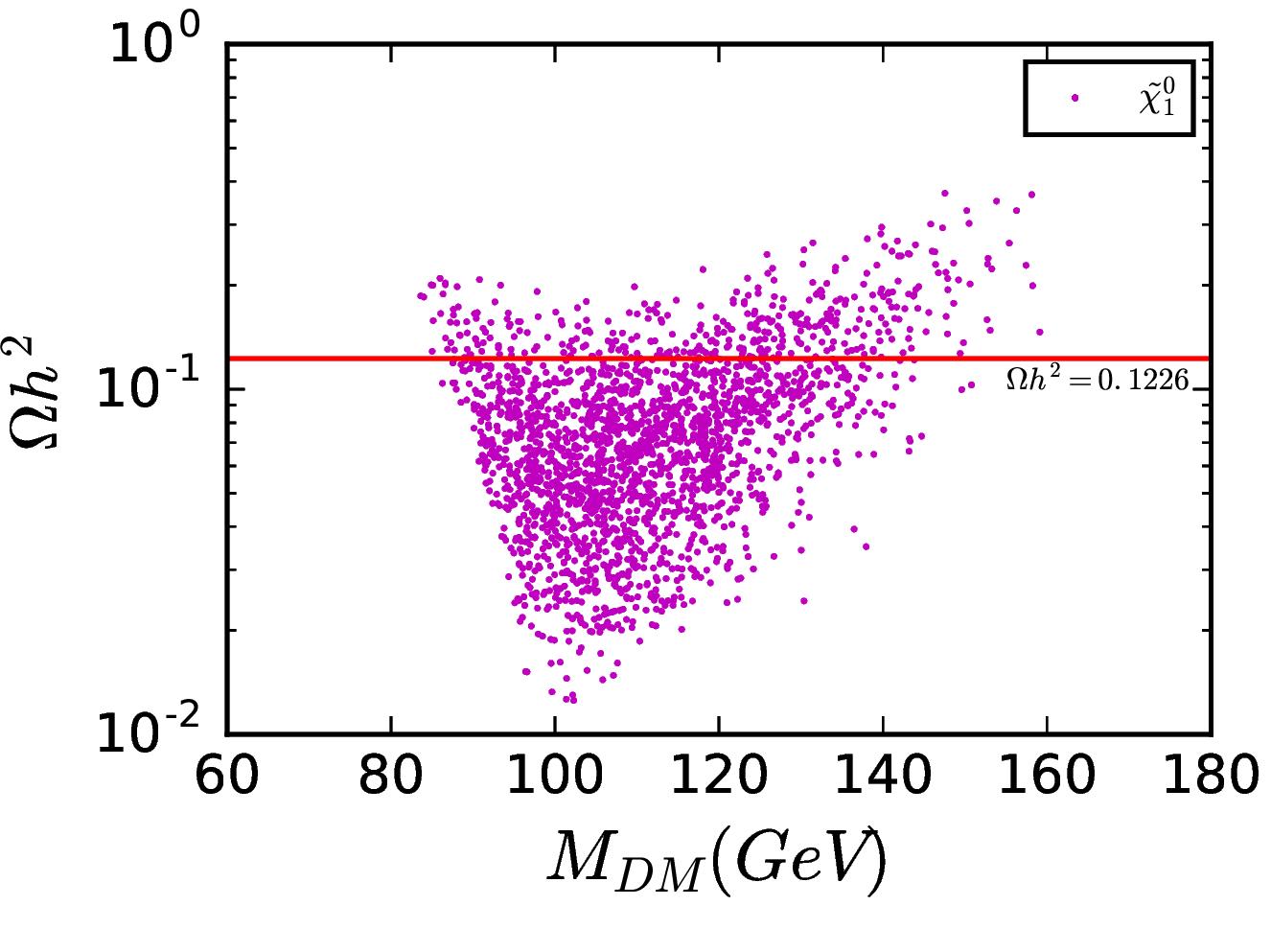}
\includegraphics[width=2.8in]{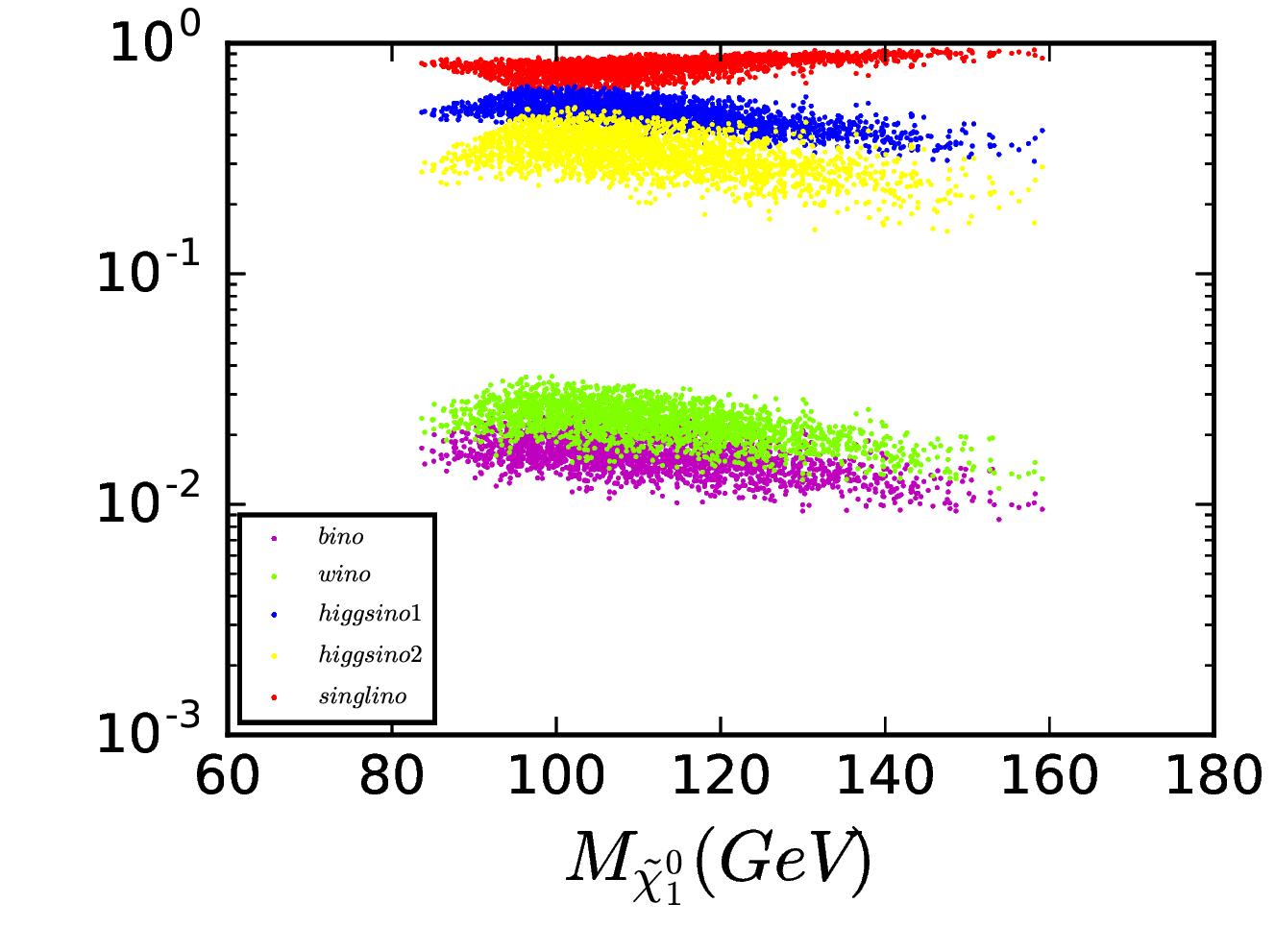}\\
\includegraphics[width=2.8in]{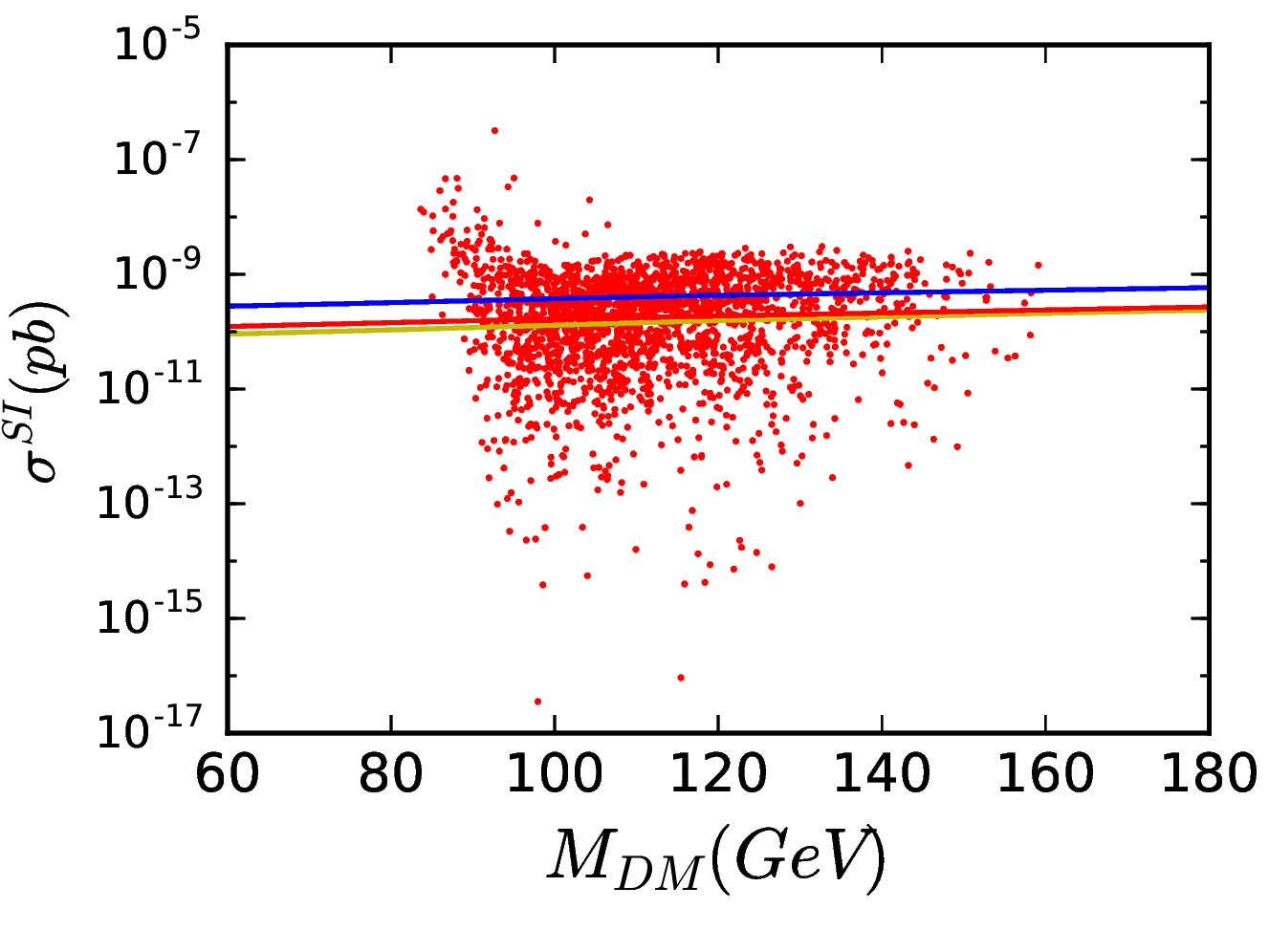}
\includegraphics[width=2.8in]{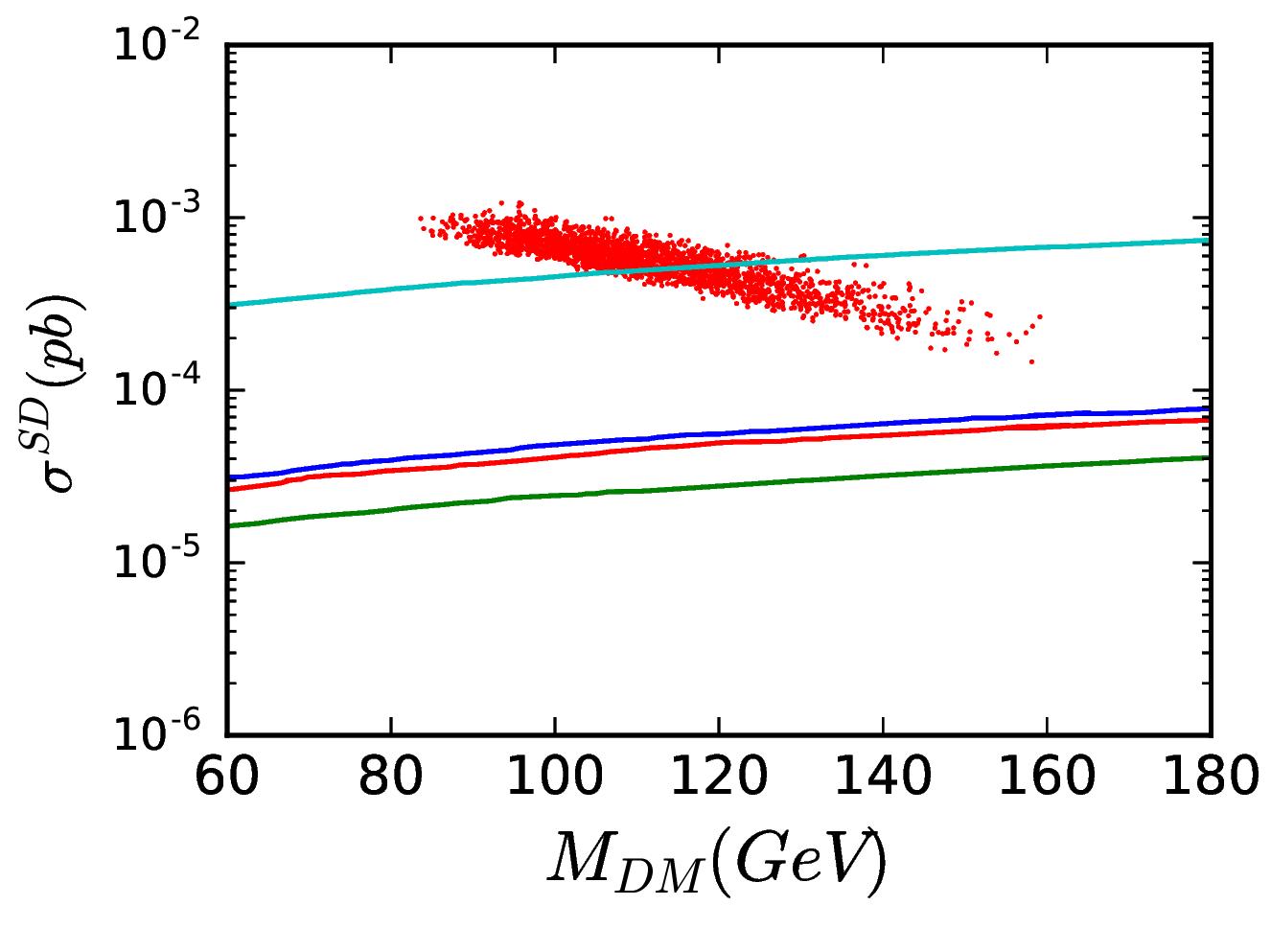}\\
\end{center}
\vspace{-.5cm}
\caption{Survived points that can satisfy the EWSB conditions and the constraints from (I) to (IV) in the case the 125 GeV Higgs is the next-to-lightest CP-even scalar in AMSB scenario. Other notations are the same as that in Fig.1 except the right panel in the second row, which shows the lightest CP-even scalar  mass $m_{h_1}$ versus the lightest CP-odd scalar mass $m_{a_1}$.}
\label{fig4}
\end{figure}

            It can be seen in the panels of fig.\ref{fig4} that the nature of SM Higgs as the next-to-lightest CP-even scalar in addition to EWSB conditions and bounds from (I) to (VI) can rule out most of the points in the parameter space. We can give similar discussions as Case A.

              Numerical results indicate that the non-trivial deflection parameter $'d'$ and couplings $\la_0,\la_1$ are absolutely necessary to obtain realistic low energy NMSSM spectrum. From the upper left panel of fig.\ref{fig4}, we can see that  the central value of $'d'$ is $-1.8$ and the couplings $\la_0,\la_1$ are constrained to take non-vanishing values. The values of $\la_1$ lie in a narrow band centered at $\la_1\approx 0.6$.
              From the upper and middle panels of fig.\ref{fig4}, we can see that the allowed $\ka$ should lie between $0.1$ to $0.15$ while $\la$ should lie between $0.21$ to $0.31$ with the iteratively obtained $\tan\beta$ lying between $6$ and $14$ for $50~{\rm TeV}\leq F_\phi\leq 130~{\rm TeV}$. We also show the possible exclusion lines from LFV, which give an upper bounds for $M_{mess}/\la_1$. The conservative requirement $M_{mess}/\la_1\lesssim (0.6\tm 10^{14}{\rm GeV})$ will set an upper bound for messenger scale to be $3.4\tm 10^{13}{\rm GeV}$.
             From the left panel in the second row of fig.\ref{fig4}, the gluino can be seen to be constrained to lie between 3.5 TeV to 6.5 TeV, which maybe accessible in the HE-LHC. It is also obvious from this panel that the 125 GeV Higgs mass can readily act as the next-to-lightest CP-even scalar. As the width of the SM-like Higgs boson is quite narrow, the masses of the lightest CP-even scalar and the lightest CP-odd scalar can not be too light so as that the 125 GeV Higgs decaying into $h_1h_1$ and $a_1a_1$ are kinetically suppressed. Otherwise, such exotic decay modes may have sizable branching ratios and in turn suppress greatly the visible signals of the SM-like Higgs boson at the LHC. We show the masses of the lightest CP-even scalar versus the lightest CP-odd scalar in the middle right panel of fig.\ref{fig4}. All the survived points can pass the constraints from the package HiggsBounds 5.3.2~\cite{HiggsBounds}. A benchmark point is shown in Table \ref{Delta2} to illustrate the typical spectrum of this scenario.
\begin{table}[htbp]
\caption{Benchmark point for type-II neutrino seesaw mechanism extension of NMSSM from AMSB in the case that the 125 GeV Higgs is the next-to-lightest CP-even scalar. All mass parameters are in the unit of GeV.}
 \begin{tabular}{|c|c|c|c|c|c|}
 \hline
 \hline
 ${F_{\phi}}$&$57060.793$&$M_{mess}$&$0.320\tm10^{14}$&$\la_0$&$0.214$\\
 \hline
$\la_1$&$0.536$&$d$&$-1.768$&$\la$&$0.254$\\
 \hline
 $\ka$&$0.120$&&&&\\ \hline\hline
 $\tan{\beta}$&$11.056$&$A_\la$&$1174.165$&$A_\ka$&$-45.938$\\
 \hline
 $A_t$&$3768.807$&$A_b$&$-243.293$&$A_\tau$&$1364.940$\\
 \hline
 $m_{h_1}$&$115.083$&$m_{h_2}$&$125.506$&$m_{h_3}$&$1389.799$\\
 \hline
 $m_{a_1}$&$91.183$&$m_{a_2}$&$1389.495$&$m_h^{\pm}$&$1390.504$\\
 \hline
 $m_{\tl{d}_{L}}$&$2893.365$&$m_{\tl{d}_{R}}$&$2552.422$&$m_{\tl{u}_{L}}$&$2892.321$\\
 \hline
 $m_{\tl{u}_{R}}$&$2584.223$&$m_{\tl{s}_{L}}$&$2893.365$&$m_{\tl{s}_{R}}$&$2552.422$\\
 \hline
$m_{\tl{c}_{L}}$&$2892.321$&$m_{\tl{c}_{R}}$&$2584.223$&$m_{\tl{b}_{1}}$&$2466.044$\\
 \hline
$m_{\tl{b}_{2}}$&$2781.567$&$m_{\tl{t}_{1}}$&$2456.970$&$m_{\tl{t}_{2}}$&$2809.467$\\
 \hline
$m_{\tl{e}_{L}}$&$1532.640$&$m_{\tl{e}_{R}}$&$917.076$&$m_{\tl{\nu_e}}$&$1530.738$\\
 \hline
$m_{\tl{\mu}_{L}}$&$1532.640$&$m_{\tl{\mu}_{R}}$&$917.076$&$m_{\tl{\nu_{\mu}}}$&$1530.738$\\
 \hline
$m_{\tl{\tau}_{1}}$&$902.423$&$m_{\tl{\tau}_{2}}$&$1528.398$&$m_{\tl{\nu_{\tau}}}$&$1526.490$\\
 \hline
$m_{\tl{\chi}_{1}^{0}}$&$112.091$&$m_{\tl{\chi}_{2}^{0}}$&$-143.316$&$m_{\tl{\chi}_{3}^{0}}$&$168.157$\\
 \hline
 $m_{\tl{\chi}_{4}^{0}}$&$-1441.722$&$m_{\tl{\chi}_{5}^{0}}$&$-2004.042$&$m_{\tl{\chi}_{1}^{\pm}}$&$141.384$\\
 \hline
 $m_{\tl{\chi}_{2}^{\pm}}$&$-2004.040$&$\mu_{eff}$&$129.966$&$m_{\tl{g}}$&$4824.773$\\
 \hline
$\Omega_\chi h^2$&$0.047$&$\sigma_P^{SI}$&$1.395\tm10^{-10}pb$&$\Delta a_\mu$&$-4.247\tm 10^{-11}$\\
 \hline
 \hline
 \end{tabular}
\label{Delta2}
\end{table}

             The lightest neutralino can be the DM candidate, which can provide full abundance of cosmic DM only in a small region. Even though  the singlino-like DM can not account for the full DM relic abundance in a large portion of the allowed parameter space, direct DM detection bounds from spin-independent cross section $\sigma_{SI}$ can rule out the majority of the survived points (see the panels in the bottom of fig.\ref{fig4}). Besides, the spin-dependent cross section $\sigma_{SD}$ can rule out the whole parameter space of this scenario.
             This can be understood from the ingredients of neutralino (shown in the right panel of the third row), in which the difference of the Higgsino components can be sizable.

    \eit

\section{\label{conclusions}Conclusions}
We propose to accommodate economically the type-II neutrino seesaw mechanism in NMSSM from GMSB and AMSB, respectively. The heavy triplets within neutrino seesaw mechanism are identified to be the messengers. Therefore, the $\mu$-problem, the neutrino mass generation, lepton-flavor-violation as well the soft SUSY breaking parameters can be economically combined in a non-trivial way. General features related to the type-II neutrino seesaw mechanism extension of NMSSM are discussed. The type-II neutrino seesaw-specific interactions can give additional Yukawa deflection contributions to the soft SUSY breaking parameters of NMSSM, which are indispensable to realize successful EWSB and accommodate the 125 GeV Higgs. Relevant numerical results, including the constraints of dark matter and possible LFV processes $l_i\ra l_j \ga$ etc, are also given. We find that our economical type-II neutrino seesaw mechanism extension of NMSSM from AMSB or GMSB can lead to realistic low energy NMSSM spectrum, both admitting the 125 GeV Higgs as the lightest CP-even scalar. The possibility of the 125 GeV Higgs being the next-to-lightest CP-even scalar in GMSB-type scenario is ruled out by the constraints from EWSB, collider and precision measurements. The possibility of the 125 GeV Higgs being the next-to-lightest CP-even scalar in AMSB-type scenario is ruled out by dark matter direct detection experiments. Possible constraints from LFV processes $l_i\ra l_j \ga$ can give an upper bound for the messenger scale.

It is interesting to distinguish between the two scenarios of type-II neutrino seesaw mechanism extension of NMSSM generated by GMSB and (d)AMSB, respectively. It is obvious from the expressions of the gaugino masses that GMSB predicts the mass ratio for gauginoes
\beqa
M_3:M_2:M_1=\al_3(M_{\rm mess}):\al_2(M_{\rm Mess}):\al_1(M_{\rm Mess}).
\eeqa
at the messenger scale $M_{\rm mess}\sim 10^{14}{\rm  GeV}$, which will lead to the approximate mass ratio
\beqa
M_3:M_2:M_1\approx 6:2:1~,
\eeqa
at the TeV scale. For our AMSB scenario, the mass ratio of gauginoes are predicted to be
\beqa
M_3:M_2:M_1=\al_3(M_{\rm mess})\(-3-7d\):\al_2(M_{\rm mess})\(1-7d\):\al_1(M_{\rm mess})\(\f{33}{5}-7d\),
\eeqa
with the values of deflection parameter $d$ centered approximately at $-1.5$ by our numerical results. So we can get the approximate mass ratio
\beqa
M_3:M_2:M_1\approx 45: 23:18.1~,
\eeqa
at TeV scale for AMSB case. If gluino can be discovered by LHC, the mass of the lightest neutralino in GMSB, whose dominant component($\gtrsim 90\%$) is the bino for most survived parameter space, can be predicted by such a gaugino mass ratio. The lightest neutralino in our AMSB scenario, on the other hand, is mostly singlino-dominant and its mass cannot be determined simply by such mass relation unless $\mu$ is known.

As noted previously, the LSP in GMSB is always the gravitino $\tl{G}$, which could act as the DM candidate. The long-lived neutralino, predicted by our scenario with $M_{mess}$ determined by heavy triplet threshold, behaves like a stable particle in the detector and its collider signatures closely resemble those of the ordinary supersymmetric scenarios with a stable neutralino. As the lightest neutralino decays into photons outside the detector, the discovery of additional high energy photon sources near the detector can be an evidence of this GMSB scenario. The AMSB scenario, however, will lead to stable neutralino. The neutralino DM of our AMSB scenario can possibly be discovered by future DM direct detection experiments, such as LUX, Xenon1T or PandaX. The gravitino DM of GMSB, which is very light, is impossible to be discovered by such experiments.

We should also brief note the differences between our scenario and ordinary deflected AMSB (GMSB). In our scenario, we need to introduce new interaction terms involving the couplings of the triplet to leptons as well as to the Higgs doublets to generate tiny neutrino masses via type-II seesaw mechanism, which will lead to new contributions to the discontinuity of the anomalous dimensions across the triplet thresholds. That is, the soft SUSY breaking parameters at the messenger scale take a different form in our scenario in contrast to that of ordinary deflected AMSB (GMSB). With these new contributions to slepton masses and $A_t$ etc, our scenarios can lead to realistic spectrum and accommodate the 125 GeV Higgs more easily than ordinary deflected AMSB (GMSB). From our numerical results, it is also clear that there is a lower bound on the scale of messenger. Such a lower bound origin from the 125 GeV Higgs and the difficulty to generate realistic NMSSM spectrum. Besides, possible LFV bounds from $l_i\ra l_j \gamma$ can set an upper bound for messenger scale. If we set the conservative requirement $M_{mess}/\la_1\lesssim (0.6\tm 10^{14}{\rm GeV})$, the messenger scale will have an upper bound to be $3.4\tm 10^{13}{\rm GeV}$ ($6.9\tm 10^{13}{\rm GeV}$) in the case of dAMSB (GMSB), respectively.
\begin{acknowledgments}
We are very grateful to the referee for helpful discussions and useful comments. This work was supported by the Natural Science Foundation of China under grant numbers 11675147,11775012.
\end{acknowledgments}
\appendix
\section{The soft SUSY breaking scalar masses from type-II neutrino seesaw mechanism extension of NMSSM}
We collect the expressions for the soft SUSY breaking scalar masses in the appendix.
\subsection{\label{appendix:A} Expressions from GMSB}
For later convenience, we list the discontinuity of various Yukawa beta functions across the messenger threshold
\beqa
\Delta \beta_{y_t}&=&\f{1}{16\pi^2}\[2\(y^u_{\bf 15}\)^2\],~~~
\Delta \beta_{y_b}=\f{1}{16\pi^2}\[2\(y^d_{\bf 15}\)^2\],~\nn\\
\Delta \beta_{\la}&=&\f{1}{16\pi^2}\[2\(y^u_{\bf 15}\)^2+2\(y^d_{\bf 15}\)^2+15\(y_S\)^2\],~~~
\Delta \beta_{\ka}=\f{1}{16\pi^2}\[45\(y_S\)^2\],~
\eeqa
and define
\beqa
\label{GMSB:DeltaG}
\Delta\tl{G}_{y_t}&\equiv&16\pi^2 \Delta \beta_{y_t}~,~~~
\Delta\tl{G}_{y_b}\equiv 16\pi^2 \Delta \beta_{y_b}~,\nn\\
\Delta\tl{G}_{\la}&\equiv&16\pi^2 \Delta \beta_{\la}~,~~~
\Delta\tl{G}_{\ka}\equiv 16\pi^2 \Delta \beta_{\ka}~.
\eeqa
The soft scalar masses are given as
\beqa
\label{GMSB:scalar}
m^2_{\tl{Q}_{L,a}}&=&\(\f{F_X^2}{M^2}\)\f{1}{(16\pi^2)^2}\[-y_t^2\Delta \tl{G}_{y_t}\delta_{a,3}-y_b^2\Delta\tl{G}_{y_b}\delta_{a,3}+\(\f{8}{3}g_3^4+\f{3}{2}g_2^4+\f{1}{30}g_1^4\)7 \],~~\nn\\
m^2_{\tl{U}^c_{L,a}}&=&\(\f{F_X^2}{M^2}\)\f{1}{(16\pi^2)^2}\[-2y_t^2\Delta \tl{G}_{y_t}\delta_{a,3}+\(\f{8}{3}g_3^4+\f{8}{15}g_1^4\)7 \],~~\nn\\
m^2_{\tl{D}^c_{L,a}}&=&\(\f{F_X^2}{M^2}\)\f{1}{(16\pi^2)^2}\[3\(y_{D D \Delta_{se};a}\)^2\tl{G}_{{D D \Delta_{se};a}}^+ + 2\(y_{L D \Delta_{3,2};a}\)^2 \tl{G}_{{L D \Delta_{3,2};a}}^+\right. \nn\\
&&~~~~~~~~~~~~~~~~~~~~~~\left.-2y_b^2\Delta \tl{G}_{y_b}\delta_{a,3}+\(\f{8}{3}g_3^4+\f{2}{15}g_1^4\)7\]~,\nn\\
m^2_{\tl{L}_{L,a}}&=&\(\f{F_X^2}{M^2}\)\f{1}{(16\pi^2)^2}\[2\(y_{L L \Delta_T;a}\)^2\tl{G}^+_{LL\Delta_T;a}
+3\(y_{L D \Delta_{3,2};a}\)^2\tl{G}^+_{{L D \Delta_{3,2};a}}\right. \nn\\
&&~~~~~~~~~~~~~~~~~~~~~~\left. +\(\f{3}{2}g_2^4+\f{3}{10}g_1^4\)7\]~,~\nn\\
m^2_{\tl{E}_{L,a}^c}&=&\(\f{F_X^2}{M^2}\)\f{1}{(16\pi^2)^2}\[\(\f{6}{5}g_1^4\)7\]~,~\nn\\
m^2_{\tl{S}}~&=&\(\f{F_X^2}{M^2}\)\f{1}{(16\pi^2)^2}\[ 3\(y_{S\overline{\Delta}_T\Delta_T}\)^2\tl{G}_{S\overline{\Delta}_T\Delta_T}^++
6\(y_{S\overline{\Delta}_{se}\Delta_{se}}\)^2\tl{G}_{S\overline{\Delta}_{se}\Delta_{se}}^+ \right.\nn\\
&&~~~~~~~~~~~~~~~~~+\left. 6\(y_{S\overline{\Delta}_{3,2}\Delta_{3,2}}\)^2 \tl{G}_{S\overline{\Delta}_{3,2}\Delta_{3,2}}^+ -2\la^2\Delta\tl{G}_\la-2\ka^2\Delta \tl{G}_{\ka}\]~,\nn\\
m^2_{H_u}~&=&\(\f{F_X^2}{M^2}\)\f{1}{(16\pi^2)^2}\[2\(y_{H_u H_u \overline{\Delta}_T}\)^2 \tl{G}_{H_u H_u \overline{\Delta}_T}^+-3y_t^2\Delta \tl{G}_{y_t}-\la^2\Delta \tl{G}_\la+\(\f{3}{2}g_2^4+\f{3}{10}g_1^4\)7\]~,\nn\\
m^2_{H_d}~&=&\(\f{F_X^2}{M^2}\)\f{1}{(16\pi^2)^2}\[2\(y_{H_d H_d {\Delta}_T}\)^2\tl{G}_{H_d H_d {\Delta}_T}^+-3y_b^2\Delta \tl{G}_{y_b}-\la^2\Delta \tl{G}_\la+\(\f{3}{2}g_2^4+\f{3}{10}g_1^4\)7\]~,\nn\\
\eeqa
with
\beqa
\tl{G}_{{D D \Delta_{se};a}}^+ &=&10 \(y^L_{\bf 15;a}\)^2+\sum\limits_{c}\(y^L_{\bf 15;c}\)^2+ \(y_X\)^2+4y_b^2\delta_{a,3}-12 g_3^2-\f{4}{3}g_1^2~,\nn\\
\tl{G}_{{L D \Delta_{3,2};a}}^+ &=&10 \(y^L_{\bf 15;a}\)^2+\sum\limits_{c}\(y^L_{\bf 15;c}\)^2+ \(y_X\)^2+2y_b^2\delta_{a,3}-\f{16}{3}g_3^2-{3}g_2^2-\f{7}{15}g_1^2~,\nn\eeqa
\beqa
\tl{G}^+_{LL\Delta_T;a}&=&10 \(y^L_{\bf 15;a}\)^2+ \(y^d_{\bf 15}\)^2+ \sum\limits_{c}\(y^L_{\bf 15;c}\)^2+ \(y_X\)^2-{7}g_2^2-\f{9}{5}g_1^2~,\nn\\
\tl{G}_{S\overline{\Delta}_T\Delta_T}^+&=&\(y^u_{\bf 15}\)^2+\(y^d_{\bf 15}\)^2+\sum\limits_{c}\(y^L_{\bf 15;c}\)^2+ 2\(y_X\)^2+2\la^2+2\ka^2-{8}g_2^2-\f{12}{5}g_1^2~,\nn\\
\tl{G}_{S\overline{\Delta}_{se}\Delta_{se}}^+&=&\sum\limits_{c}\(y^L_{\bf 15;c}\)^2+ 2\(y_X\)^2+2\la^2+2\ka^2-\f{40}{3}g_3^2-\f{16}{15}g_1^2~,\nn\\
\tl{G}_{S\overline{\Delta}_{3,2}\Delta_{3,2}}^+&=&\sum\limits_{c}\(y^L_{\bf 15;c}\)^2+ 2\(y_X\)^2+2\la^2+2\ka^2-\f{16}{3}g_3^2-{3}g_2^2-\f{1}{15}g_1^2~,\nn\\
\tl{G}_{H_u H_u \overline{\Delta}_T}^+&=&3\(y^u_{\bf 15}\)^2+ \(y_X\)^2+6y_t^2-{7}g_2^2-\f{9}{5}g_1^2~,\nn\\
\tl{G}_{H_d H_d {\Delta}_T}^+&=&\sum\limits_{c}\(y^L_{\bf 15;c}\)^2+3\(y^d_{\bf 15}\)^2+ \(y_X\)^2+6y_b^2-{7}g_2^2-\f{9}{5}g_1^2~,
\eeqa
and
\beqa
&&y_{S\overline{\Delta}_T\Delta_T}=y_{S\overline{\Delta}_{se}\Delta_{se}}
=y_{S\overline{\Delta}_{3,2}\Delta_{3,2}}=y_S~,\nn\\
&&y_{L L \Delta_T;a}=y_{D D \Delta_{se};a}=y_{L D \Delta_{3,2};a}=y^L_{\bf 15;a}~,\nn\\
&&y_{H_u H_u \overline{\Delta}_T}=y^u_{\bf 15}~,~~~~~~~y_{H_d H_d {\Delta}_T}=y^d_{\bf 15}~.
\eeqa
\subsection{\label{appendix:B} Expressions from deflected AMSB}
The expressions of $\delta_G$ can be obtained by the following replacement
\beqa
\f{F_X}{M}\ra d F_\phi~,~~~y_S=0~,
\eeqa
in eqn(\ref{GMSB:scalar}). The expressions of $\delta_{A}$ are given by ordinary AMSB predictions
\beqa
\delta^A_{{H}_u}~~&=&\f{F_\phi^2}{16\pi^2}\[\f{3}{2}G_2\al^2_2+\f{3}{10}G_1\al^2_1\]
+\f{F_\phi^2}{(16\pi^2)^2}\[\la^2\tl{G}_\la+3y_t^2\tl{G}_{y_t}\]~,~\nn\\
\delta^A_{{H}_d}~~&=&\f{F_\phi^2}{16\pi^2}\[\f{3}{2}G_2\al^2_2+\f{3}{10}G_1\al^2_1\]
+\f{F_\phi^2}{(16\pi^2)^2}\[\la^2\tl{G}_\la+3y_b^2\tl{G}_{y_b}\]~,~\nn\\
\delta^A_{\tl{Q}_{L;a}}&=&\f{F_\phi^2}{16\pi^2}\[\f{8}{3} G_3 \al^2_3+\f{3}{2}G_2\al^2_2+\f{1}{30}G_1\al^2_1\]
+\delta_{a,3}\f{F_\phi^2}{(16\pi^2)^2}\[y_t^2\tl{G}_{y_t}+y_b^2\tl{G}_{y_b}\]~,~\nn\\
\delta^A_{\tl{U}^c_{L;a}}&=&\f{F_\phi^2}{16\pi^2}\[\f{8}{3} G_3 \al^2_3+\f{8}{15}G_1\al^2_1\]+\delta_{a,3}\f{F_\phi^2}{(16\pi^2)^2}\[2y_t^2\tl{G}_{y_t}\]~,~\nn\\
\delta^A_{\tl{D}^c_{L;a}}&=&\f{F_\phi^2}{16\pi^2}\[\f{8}{3} G_3 \al^2_3+\f{2}{15}G_1\al^2_1\]+\delta_{a,3}\f{F_\phi^2}{(16\pi^2)^2}\[2y_b^2\tl{G}_{y_b}\]~,~\nn\\
\delta^A_{\tl{L}_{L;a}}&=&\f{F_\phi^2}{16\pi^2}\[\f{3}{2}G_2\al_2^2+\f{3}{10}G_1\al_1^2\]~,~\nn\\
\delta^A_{\tl{E}_{L;a}^c}&=&\f{F_\phi^2}{16\pi^2}\f{6}{5}G_1\al_1^2~,~\nn\\
\delta^A_{S}&=&\f{F_\phi^2}{(16\pi^2)^2}\[2\la^2\tl{G}_{\la}+2\ka^2\tl{G}_{\ka}\]~,
\eeqa
with
\beqa
G_i&=&-b_i~,~~~~~~~~~~~~~~~~(b_1,b_2,b_3)=(\f{33}{5},1,-3)~.
\eeqa
The expressions of the gauge-anomaly interference terms are
\beqa
2\delta^I_{\tl{Q}_{L,a}}&=&-\f{d F_\phi^2}{(8\pi^2)^2}\[\delta_{a,3} y_t^2\Delta \tl{G}_{y_t}+\delta_{a,3} y_b^2\Delta\tl{G}_{y_b}-7\(\f{8}{3}g_3^4+\f{3}{2}g_2^4+\f{1}{30}g_1^4\)\]~,\nn\\
~~2\delta^I_{\tl{U}^c_{L,a}}&=&-\f{d F_\phi^2}{(8\pi^2)^2}\[\delta_{a,3}2y_t^2\Delta \tl{G}_{y_t}-7\(\f{8}{3}g_3^4+\f{8}{15}g_1^4\)\]~,\nn\\
~~2\delta^I_{\tl{D}^c_{L,a}}&=&-\f{d F_\phi^2}{(8\pi^2)^2}\[\delta_{a,3}2y_b^2\Delta \tl{G}_{y_b}-7\(\f{8}{3}g_3^4+\f{2}{15}g_1^4\)\]~,\nn\\
~~2\delta^I_{\tl{L}_{L,a}}&=&-\f{d F_\phi^2}{(8\pi^2)^2}\[-7\(\f{3}{2}g_2^4+\f{3}{10}g_1^4\)\]~,\nn\\
~~2\delta^I_{\tl{E}^c_{L,a}}&=&-\f{d F_\phi^2}{(8\pi^2)^2}\[-7\(\f{6}{5}g_1^4\)\]~,\nn\\
2\delta^I_{{H}_u}&=&
-\f{d F_\phi^2}{(8\pi^2)^2}\[\la^2\Delta \tl{G}_\la+3y_t^2\Delta \tl{G}_{y_t}-7\(\f{3}{2}g_2^4+\f{3}{10}g_1^4\)\]~,\nn\\
2\delta^I_{{H}_d}&=&-\f{d F_\phi^2}{(8\pi^2)^2}\[\la^2\Delta\tl{G}_\la+3y_b^2\Delta \tl{G}_{y_b}-7\(\f{3}{2}g_2^4+\f{3}{10}g_1^4\)\]~,~\nn\\
2\delta^I_{S}&=&-\f{d F_\phi^2}{(8\pi^2)^2}\(2\la^2\Delta\tl{G}_\la+2\ka^2\Delta \tl{G}_{\ka}\)~,~
\eeqa
with $\Delta \tl{G}_{y_b},\Delta \tl{G}_{y_b},\Delta\tl{G}_\la, \Delta \tl{G}_{\ka}$ given in
eqn (\ref{GMSB:DeltaG}).

\end{document}